%% file: main.tex
\documentclass[%
reprint,
superscriptaddress,
groupedaddress,
amsmath,amssymb,
aps,
pra,
]{revtex4-2}

\usepackage{graphicx}
\usepackage{dcolumn}
\usepackage{bm}
\usepackage{xcolor}
\usepackage{qcircuit}
\usepackage{braket}
\usepackage{dsfont}
\usepackage{mathrsfs}
\usepackage{placeins}
\usepackage{comment}
\usepackage[utf8]{inputenc}
\usepackage{pgfplots}
\usepackage{hyperref}
\usepackage{cleveref}
\usepackage{cleveref}
\usepackage{array}
\usepackage{tabularx}
\usepackage{booktabs}
\usepackage{pgfplots}

\DeclareUnicodeCharacter{2212}{−}
\usepgfplotslibrary{groupplots,dateplot}
\usetikzlibrary{patterns,shapes.arrows}
\pgfplotsset{compat=newest}

\begin{document}

\title{Analytic leakage suppression with a single control field: fast two-qubit gates with tunable couplers}

\author{Lukas Heunisch\textsuperscript{1, 2, 3}}
\email{lukas.heunisch@fau.de}
\author{Michael J. Hartmann\textsuperscript{1, 3}}
\author{Aashish A. Clerk\textsuperscript{2}}
\email{aaclerk@uchicago.edu}

\affiliation{\textsuperscript{1}Physics Department, Friedrich-Alexander-Universität Erlangen-Nürnberg, Germany}
\affiliation{\textsuperscript{2}Pritzker School of Molecular Engineering and Chicago Quantum Institute,
The University of Chicago, 60637 Chicago, Illinois, USA}
\affiliation{\textsuperscript{3}Quint Computing GmbH, 91058 Erlangen, Germany}

\date{\today}

\begin{abstract}
Simple analytic pulse-shaping techniques are of great practical utility in quantum control, with prime examples being the DRAG method for suppressing leakage in superconducting microwave gates and the transitionless-driving approach to shortcuts-to-adiabaticity. Standard versions of these methods require two orthogonal control channels, with the second channel effectively breaking time-reversal symmetry. This appears to rule out their use in settings with only a single real-valued control field, such as
the kind of baseband flux control that is common in many superconducting circuit architectures. We show here that a simple analytic pulse-shaping technique derived via a Magnus expansion is effective even with just a single baseband control channel. We demonstrate its efficacy by simulating a two-qubit gate between transmons realized with a tunable coupler and baseband flux pulses.  Our corrections dramatically reduce non-adiabatic leakage caused by ramping the coupler: for realistic device parameters, leakage in a fast iSWAP gate is suppressed by up to three orders of magnitude. Our approach is general, goes beyond simply suppressing unwanted spectral weight at leakage transitions, and can be applied to a variety of platforms.    
\end{abstract}

\maketitle

\input{./1_Introduction}
\input{2_iSWAP_AD_frame}
\input{3_Leakage_correction_in2exc}
\input{4_Results}
\input{5_Slepian}
\input{6_Conclusion}

\appendix
\input{./Appendix.tex}

\FloatBarrier

\nocite{*}
\bibliography{mybib}

\end{document}

%% file: 1_Introduction.tex
\section{Introduction}

Fast, high-fidelity two-qubit gates are a central requirement for quantum information processing. While numerical optimal control provides a powerful route for their design 
(see e.g.~\cite{Werninghaus2021npj, Glaser2015EPJD, Koch2022EPJD, Machnes2018PRL, Kelly2014PRL}), 
there remains practical value in simple, analytic pulse-shaping techniques: they are transparent, easy to calibrate, and can reveal the physics underlying the relevant errors. The paradigmatic example is the DRAG protocol \cite{Motzoi2009PRL, Gambetta2011PRA, Motzoi2013PRA, Theis2018EPL}, which suppresses leakage to non-computational states in weakly anharmonic qubits such as transmons, and which is by now a standard tool for single-qubit gates in superconducting circuits \cite{Lucero2010PRA, Chow2010PRA}. 
A related strategy is the ``transitionless driving" (TD) approach to shortcuts to adiabaticity, which also yields a simple analytic route to suppressing non-adiabatic errors \cite{Demirplak2003JPCA,Berry2009JPAM}.  
Suppressing leakage is especially important in quantum error correction (QEC) settings: standard codes are designed to correct Pauli errors within the computational subspace and do not natively mitigate leakage \cite{Aliferis2007QIC, Suchara2015QIC, Google2024Nature, Krinner2022Nature}.  

The best-known versions of the above analytic control techniques share a crucial structural requirement: two independent, orthogonal control channels, with the second channel effectively breaking time-reversal symmetry. In DRAG, these are the real-valued in-phase and quadrature-phase amplitudes of a resonant microwave drive, providing independent $\sigma_x$ and $\sigma_y$.  The basic TD protocol also requires independent  $\sigma_x$ and $\sigma_y$ controls. 
This requirement would seem to exclude a large and technologically important class of gates based on a single control channel. An important example is in modern superconducting circuits:  rapid two-qubit operations here increasingly rely on tunable couplers, activated by baseband flux pulses that tune the coupler frequency (see 
e.g.~\cite{Yan2018PRApplied, Li2024PRX, Collodo2020PRL, Sung2021PRX, Sete2021PRApplied, Stehlik2021PRL, Foxen2020PRL}). 
These gates are controlled via a single time-dependent flux:  there is no carrier, no phase, and hence no second quadrature. At the same time, the fast ramping of the coupler required for gate speed produces non-adiabatic leakage out of the computational subspace; this is often the dominant coherent error. The lack of a second control channel here seems to imply that DRAG or TD analytic leakage cancellation is
precluded.  Mitigating leakage in these gates has thus largely relied on brute-force numerical optimization (see e.g. \cite{Werninghaus2021npj, Krauss2026arxiv, Sarma2025PRApplied}).

In this work, we show that effective analytic leakage suppression is in fact possible in this more restricted setting of a single baseband control field.  
We start with the realization that both DRAG and TD function by approximately canceling non-adiabatic transitions {\it on average} during the duration of the gate, with the system deviating from the ideal evolution at intermediate times (see Fig. \ref{fig:Scheme}c).  The idea is to formulate this condition as broadly as possible, to accommodate the widest possible set of underlying control.  This can effectively be done using a Magnus expansion, building on the general strategy of Ref.~\cite{Ribeiro2017PRX}.  Doing this, one finds that DRAG and TD are just two specific (and constrained) solutions to this problem, whereas other modified strategies, as we present, are fully compatible with baseband control pulses.    

We demonstrate the method on a setting ubiquitous in superconducting circuits: an iSWAP gate between transmon qubits mediated by a tunable coupler and actuated by baseband flux pulses.  Our simulations show that for realistic device parameters, our analytic correction method suppresses leakage by up to three orders of magnitude.  We also show how our strategy can be used to mitigate simultaneous leakage channels, applied to other kinds of two-qubit gates, and made even more flexible by working in higher adiabatic frames.  While our focus is on superconducting circuits, the basic strategy could be employed in any setting where gates are driven by a single baseband control parameter, e.g.~voltage control of exchange interactions in semiconductor spin qubits \cite{Loss1998PRA, Polat2025arxiv, Burkard2023RMP}. 

The rest of the paper is structured as follows. In 
Sec.~\ref{sec:single_exc_subspace}, we give a basic example of our approach, showing how it can suppress single-excitation leakage to a coupler in an iSWAP gate.
In Sec.~\ref{sec:leak_2subspace}, we extend this strategy to correct leakage errors in this gate occurring in the two-excitation subspace.  We demonstrate the effectiveness of our corrected analytic pulses in Sec.~\ref{sec:Hardware_result} by simulating an iSWAP gate on a realistic hardware Hamiltonian describing two transmons coupled by a tunable transmon coupler.  In Sec.~\ref{sec:two_channel_corr}, we explore how our methods can be extended to simultaneously address multiple leakage pathways, and in Sec.~\ref{sec:Slepian},  we show how it can be combined with existing optimal-pulse methods (e.g.~with optimized Slepian-based pulse shapes as used recently in the experiment of \cite{Sung2021PRX}).  We end with a brief summary and outlook in 
Sec.~\ref{sec:Conclusions}.

Finally note that during the writing of this manuscript we became aware of Ref.~\cite{Georgiadis2026arxiv}, which also derives correction pulses for flux-pulse activated two-qubit gates. Their approach is distinct from the one presented here. Ref.~\cite{Georgiadis2026arxiv} cancels the spectral weight of the effective coupling at the leakage transition frequency, whereas our Magnus-based strategy does not rely on a Fourier picture and remains applicable when this frequency varies during the gate. A detailed comparison of both methods is given in Appendix~\ref{sec:PhiDRAG}.

%% file: 2_iSWAP_AD_frame.tex
\section{Leakage correction in the single-excitation subspace}
\label{sec:LeakageSingleMain}
While we will be interested in more complex multi-level settings, to illustrate our basic ideas, we start our analysis with the simplest non-trivial setting: the single-excitation Hamiltonian of a qubit--tunable coupler--qubit architecture (see Fig. \ref{fig:Scheme}(a)), motivated by recent experiments \cite{Collodo2020PRL, Sung2021PRX, Sete2021PRApplied, Stehlik2021PRL, Foxen2020PRL}. Although our strategy is broadly applicable to state-of-the-art optimized pulse shapes (see e.g. Section~\ref{sec:Slepian}) and to a wide class of gates using baseband pulses (see Appendix~\ref{sec:CZ_correction} for another example), we demonstrate it here using a concrete example. The goal will be to perform an iSWAP gate between the two qubits using time-dependent flux control, in a way that mitigates non-adiabatic leakage errors stemming from the finite gate speed.

\subsection{iSWAP gate in the adiabatic frame}
\label{sec:single_exc_subspace}
\begin{figure}
    \centering
    \includegraphics[width=\columnwidth]{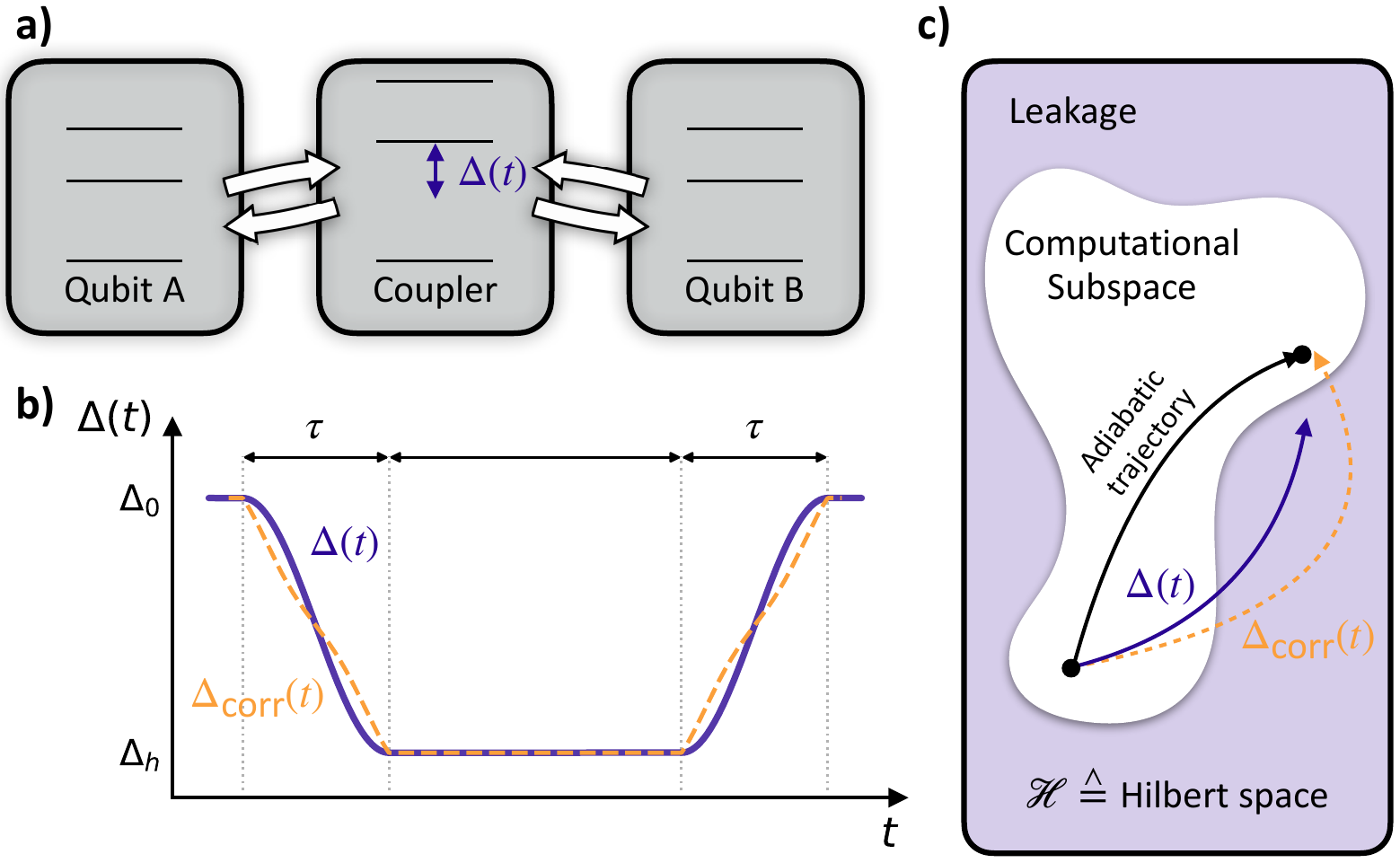}
    \caption{a) Schematic of a generic qubit--tunable coupler--qubit architecture. b) Time-dependent coupler detuning $\Delta(t)$ (realized via baseband flux control), ramping from the idle detuning $\Delta_0$ to the hold value $\Delta_h$ for a finite gate duration $t_g$.
    The goal is to start with a base pulse shape (blue) and modify it slightly (orange) to mitigate leakage errors. c) Schematic of relevant trajectories in the full Hilbert space $\mathscr{H}$.
    The goal is an ideal evolution between computational states (white).  A finite gate time leads to non-adiabatic errors, and population of leakage states (purple, blue trajectory).  The correction strategy is to use available controls to cancel leakage at the final time, not at every instant during the gate (orange trajectory).  }
    \label{fig:Scheme}
\end{figure}

  Two fixed-frequency, resonant qubits exchange excitations via a central tunable coupling element, which is modeled as a frequency-tunable qubit. Expressed in the basis $\{\ket{1}, \ket{c}, \ket{2}\}$ and in a frame rotating at the bare qubit frequency $\omega_1 = \omega_2 = \omega$, the Hamiltonian reads
\begin{equation}
\label{eqn:H_lab}
    H(t) = \begin{pmatrix}
        0 & g & g_{12} \\
        g & \Delta(t) & g \\
        g_{12} & g & 0 
    \end{pmatrix} \ .
\end{equation}
Here, $g$ is the qubit--coupler tunneling, which is assumed equal for both qubits, and $\Delta(t) = \omega_c(t) - \omega$ is the time-dependent coupler--qubit detuning. 
We also include a direct qubit--qubit coupling $g_{12} \ll g$,
which is parametrically small but non-zero in most 
implementations \cite{Sung2021PRX, Collodo2020PRL, Yan2018PRApplied, Heunisch2023PRApplied}. 

The gate relies on tuning $\Delta(t)$ from an initially large value
$\Delta_0$, where the qubits are effectively decoupled, to a smaller value $\Delta_h$ where there is an appreciable coupler-mediated qubit--qubit interaction $\sim g^2/\Delta_h$.  The detuning is held at this value to perform the desired gate operation, after which it is ramped back to its large off-position $\Delta_0$ (cf. Fig. \ref{fig:Scheme} b)).

At this level of approximation, the only sources of gate infidelity are coherent non-adiabatic errors generated by the time-dependence of $\Delta$.  These could lead to an unwanted non-zero coupler population at the end of the gate, i.e.~a leakage error; our goal is to describe and mitigate this process.  The standard approach for describing this kind of gate is perturbative, based on a time-dependent Schrieffer--Wolff transformation (see e.g.~\cite{Yan2018PRApplied}).  We instead take an alternate approach that makes non-adiabatic errors more transparent.  We work in the adiabatic frame, where the instantaneous eigenstates of $\hat{H}(t)$ become stationary. In this frame, the ideal gate corresponds to perfect adiabatic evolution with carefully tuned dynamical phases.  

The gate dynamics in the adiabatic frame has some analogy to adiabatic analysis of a $\Lambda$ system in the context of a STIRAP protocol \cite{Bergmann1998RMP, Vitanov2017RMP}.  While our gate is distinct from STIRAP, it also makes use of an adiabatic eigenstate whose energy is time-independent, the so-called ``dark state" \cite{Sung2021PRX}:  
  % Despite this distinction, the structural analogy remains valid, as one of the instantaneous eigenstates of our Hamiltonian in Eq. \ref{eqn:H_lab} is a decoupled ``dark state''
\begin{equation}
    \ket{D} = \frac{1}{\sqrt{2}}\bigl(\ket{1} - \ket{2}\bigr) .
\end{equation}
This state has no coupler component, is thus independent of $\Delta(t)$ and has a fixed energy $E_D = -g_{12}$. 

The remaining two instantaneous eigenstates of $\hat H(t)$ are bright states: a lower-energy qubit-like state $\ket{B_q(\theta)}$ and a higher-energy coupler-like state $\ket{B_c(\theta)}$.
Defining the time-dependent mixing angle $\theta(t)$ via
\begin{equation}
    \label{eq:theta}
    \tan (2\theta (t)) = \frac{2\sqrt{2} g}{g_{12} - \Delta(t)} \ ,
\end{equation}
the bright states are given by
\begin{align}
    \ket{B_q(\theta)} &= \phantom{-} \frac{1}{\sqrt{2}} \cos \theta \bigl( \ket{1} + \ket{2} \bigr) + \sin \theta \ket{c}  \notag \\ 
    \ket{B_c(\theta)} &= - \frac{1}{\sqrt{2}} \sin \theta \bigl( \ket{1} + \ket{2} \bigr) + \cos \theta \ket{c} \ ,
    \label{eqn:BqBc}
\end{align}
with corresponding eigenenergies
\begin{align}
\label{eqn:Eq}
    E_q &= \frac{1}{2} \bigl( g_{12}+\Delta - \sqrt{(g_{12} - \Delta)^2  + 8g^2} \bigr) \\
    E_c &= \frac{1}{2} \bigl( g_{12}+\Delta + \sqrt{(g_{12} - \Delta)^2  + 8g^2} \bigr) \ .
\label{eqn:Ec}
\end{align}

Consider now the ideal gate dynamics in terms of the instantaneous eigenstates.  We start at $t=0$ in the off-configuration $\Delta(0) = \Delta_0$, which we define by 
\begin{equation}
\label{eqn:Delta_0}
    \Delta_0 = -g_{12} + \frac{g^2}{g_{12}} \ .
\end{equation}

For this detuning, the qubit-like bright state $|B_q \rangle$
is degenerate with the dark state (see Appendix \ref{sec:coupler_off} for further details). This defines a degenerate subspace of dressed qubit states:
\begin{align}
\label{eqn:tilde1}
    \ket{\tilde{1}} &= \frac{1}{\sqrt{2}}\bigl(\ket{B_q(\theta_0)} + \ket{D}\bigr) \approx \ket{1} + \frac{\theta_0}{\sqrt 2} \ket{c} + \mathcal{O}(\theta_0^2)\\
    \ket{\tilde{2}} &= \frac{1}{\sqrt{2}}\bigl(\ket{B_q(\theta_0)} - \ket{D}\bigr) \approx \ket{2} + \frac{\theta_0}{\sqrt 2} \ket{c} + \mathcal{O}(\theta_0^2) \ ,
    \label{eqn:tilde2}
\end{align}
where $\theta_0$ is determined by Eq.~(\ref{eq:theta}) for $\Delta(t) \rightarrow \Delta_0$. 
As promised, for this detuning the coupler is effectively off, i.e.~it only weakly dresses the qubit states, but leaves them degenerate. 

We next consider the gate dynamics over an interval $t \in (0,t_g)$.  We consider a trajectory $\Delta(t)$ such that
$\Delta(0) = \Delta(t_g) = \Delta_0$ (i.e.~the qubit--qubit interaction starts off, and ends off).
In the middle of the gate, $\Delta(t)$ is ramped to and held at a value $\Delta_h < \Delta_0$, effectively turning on an interaction.  For perfect adiabatic evolution, the bright state 
$\ket{B_q(\theta)}$ accumulates a relative dynamical phase with respect to the dark state $\ket{D}$. To achieve a full SWAP rotation, this phase should satisfy
\begin{equation}
    \pi \overset{!}{=} \left| \int_0^{t_{g}} \bigl(E_q(t) - E_D\bigr) \, dt \right| \ .
    \label{eqn:gate_cond}
\end{equation}
It follows directly from 
Eqs.~(\ref{eqn:tilde1}) and (\ref{eqn:tilde2}) that such a dynamical phase swaps the qubit states $\ket{\tilde{1}}$ and $\ket{\tilde{2}}$. 

To now understand non-adiabatic errors, we rigorously transform to the adiabatic frame (in which instantaneous eigenstates become stationary).  We define
\begin{equation}
     \ket{\psi_{\text{ad}}(t)} = \hat U^\dagger_{\rm ad}(t) \ket{\psi_{\text{lab}}(t)}.
\end{equation}
Ordering the instantaneous eigenstates as $\{ \ket{B_c}, \ket{B_q}, \ket{D}\}$, the unitary transformation associated with this frame change is
\begin{equation}
    U_{\rm ad}(t) = \frac{1}{\sqrt{2}} \begin{pmatrix} 
        -\sin\theta(t) & \cos\theta(t) & 1 \\ 
        \sqrt{2}\cos\theta(t) & \sqrt{2}\sin\theta(t) & 0 \\ 
        -\sin\theta(t) & \cos\theta(t) & -1 
    \end{pmatrix} .
    \label{eqn:U_matrix}
\end{equation} 
Since the dark state completely decouples from the dynamics, the non-trivial evolution in the adiabatic frame 
is described by a $2\times2$ Hamiltonian acting on the bright states:
\begin{align}
    \hat H_{\text{ad}}(t) &= \frac{E_c(t) - E_q(t)}{2} (\ket{B_c}\bra{B_c}-\ket{B_q}\bra{B_q}) \notag \\
    &+ i\dot{\theta} (\ket{B_q}\bra{B_c}-\ket{B_c}\bra{B_q}) \ .
    \label{eqn:H_ad_2x2}
\end{align}
The last term $\propto \dot{\theta}$ is directly responsible for non-adiabatic errors.  It is only appreciable during the turn-on and turn-off period of the gate, and drives population to the coupler-like bright state. It thus results in leakage errors: at the end of the gate, population will be transferred out of the computational space spanned by $\ket{\tilde{1}},\ket{\tilde{2}}$.

\subsection{Leakage suppression via simple pulse shaping}
\label{sec:leak_corr}

The simplest and most direct way to  suppress leakage, while still having a finite gate time $t_g$ would be to directly cancel the last term in Eq.~(\ref{eqn:H_ad_2x2}).  This is the most basic kind of shortcuts-to-adiabaticity strategy
\cite{GueryOdelin2019RMP}: one adds extra time-dependent terms to the Hamiltonian such that in the adiabatic frame defined by Eq.~(\ref{eqn:U_matrix}), the $\dot{\theta}$ is canceled.  This strategy is known both as  counteradiabatic driving \cite{Demirplak2003JPCA}
and transitionless driving \cite{Berry2009JPAM}.  It suffers from a general problem: the needed additional Hamiltonian terms must effectively break time-reversal symmetry (i.e.~they are imaginary)
\cite{Berry2009JPAM, Baksic2016PRL, Sels2017PNAS}, 
something that cannot be achieved simply by changing the time-dependence of $\Delta(t)$.  Note that the situation is very different from the standard scenario, where DRAG is used \cite{Motzoi2009PRL}.  There, one is applying an AC microwave drive centered at a non-zero carrier frequency.  The needed time-reversal breaking field is easily achievable, as one controls both quadratures of the drive amplitude.  

At first glance, it would seem that our setting (based on baseband control) is incompatible with a DRAG-type strategy, as there is no simple way to introduce an orthogonal control field.  We now show that this conclusion is too pessimistic.  While an exact cancellation of the deleterious $\dot{\theta}$ term might not be possible, we can instead try to cancel it {\it on average} over the gate evolution by {\it only} modifying the shape of the real-valued detuning $\Delta(t)$, i.e.
\begin{align}
    \Delta (t) \rightarrow \Delta(t) + \delta \omega (t) \equiv \Delta_{\text{corr}}(t) \ .
\end{align}
We will do this by adapting the general Magnus-expansion-based strategy of  Ref.~\cite{Ribeiro2017PRX}.

To understand how this is possible, note that in the adiabatic frame the modification of $\Delta(t)$ yields new Hamiltonian terms: 
\begin{align}
    H_{\text{ad}}(t) & \rightarrow H_{\text{ad}}'(t)  = H_{\text{ad}}(t) + H_{\text{ctrl,ad}}(t) \\
    \hat H_{\text{ctrl,ad}}(t) &= \delta \omega(t) \Big( \cos^ 2\theta \, \ket{B_c}\bra{B_c} + \sin^2\theta \, \ket{B_q}\bra{B_q} \notag \\
    &+ \frac{\sin 2\theta}{2} \big( \ket{B_c}\bra{B_q} + \ket{B_q}\bra{B_c} \big) \Big) \ .
\label{eqn:H_ad}
\end{align}
Crucially,  the new $\delta \omega(t)$ terms also couple the two bright states.  To see whether a cancellation with the deleterious $\dot{\theta}$ terms is possible, 
it is helpful to make a further interaction-picture transformation 
generated by 
\begin{align}
    \hat H_0(t) = \frac{E_c-E_q}{2}\big(\ket{B_c}\bra{B_c} - \ket{B_q}\bra{B_q}\big) \ .
\end{align}
Defining 
$\hat U_I(t) = \exp\left(-i \int_0^t \hat H_0(t_1) \, dt_1 \right)$, the interaction-picture Hamiltonian
is $H_{\text{ad}, I}' = U_I(t)^\dagger H_{\text{ad}}' U_I(t) - H_0(t)$.
It takes the explicit form:
\begin{align}
\hat H'_{\text{ad},I}(t) &= \delta \omega(t) \Big( \cos^ 2\theta \, \ket{B_c}\bra{B_c} + \sin^2\theta \, \ket{B_q}\bra{B_q} \Big) \notag \\
&+ \frac{\sin 2\theta}{2}\delta \omega (t) \big( \ket{B_c}\bra{B_q}e^{i\Phi (t)} + \text{H.c.} \big) \notag \\
&- i\dot{\theta} \big(\ket{B_c}\bra{B_q}e^{i\Phi(t)}-\text{H.c.}\big)
\end{align}
where
\begin{equation}
    \Phi(t) = \int_0^t \bigl(E_c - E_q \bigr) \, dt_1 
\end{equation}
is a dynamical phase. 
In this frame, the ideal gate dynamics corresponds to the identity; any non-trivial dynamics thus represents an error.  
Our goal thus translates into finding a $\delta \omega(t)$ such that the interaction-picture propagator is the identity at the end of the gate.  

We next use a Magnus expansion to approximate the time-evolution operator $\hat {\mathcal{U}}(t)$ generated by this 
Hamiltonian:
\begin{gather}
    \hat {\mathcal{U}}(t) = \exp\bigl(\hat \Omega_1(t) + \hat \Omega_2(t) + \dots\bigr)  \\
    \hat \Omega_1(t) = -i\int_0^{t} \hat H_{\text{ad},I}'(t_1) \, dt_1  \\ 
    \hat \Omega_2(t) = -\frac{1}{2} \int_0^{t} dt_1 \int_{0}^{t_1} dt_2 \, [\hat H_{\text{ad}, I}'(t_1), \hat H_{\text{ad}, I}'(t_2)] \ .
\end{gather}

We can now approximately (to leading order in the Magnus expansion) attempt to enforce the condition that our control corrections cancel the unwanted leakage processes at the end of the gate at $t=t_g$. We do this by insisting that $\Omega_1(t_g)$ does not couple the two bright states,
\begin{align}
\label{eqn:Null_cond}
    0 &\overset{!}{=} \int_0^{t_g} \langle B_c | \hat H_{\text{ad},I}'(t) | B_q \rangle \, dt \\
    &= \int_0^{t_g} \left( \frac{\delta\omega(t)}{2}\sin2\theta - i\dot{\theta} \right) e^{i\Phi(t)} \, dt \ .
\end{align}

At this stage, we seem to have run into the same issue we started with: the two terms in the integrand (corresponding to the control correction and the non-adiabatic error) are always out of phase by $\pi/2$, meaning that they can never cancel instantaneously. We however only need the total integral to vanish. To achieve this, we use integration by parts to make the non-adiabatic error term in-phase with the control correction term. Assuming that our detuning profile $\Delta(t)$ is sufficiently smooth (i.e.~its first derivative vanishes smoothly at $t=0,t_g$), we have:
\begin{align}
    -i &\int_0^{t_{g}} dt \ \dot \theta e^{i\Phi(t)} \\
    &= -i\int_0^{t_{g}} dt \ \frac{\dot \theta}{i(E_c-E_q)} \frac{d}{dt} e^{i\Phi(t)} \\
    &= -\underbrace{\frac{\dot \theta e^{i\Phi(t)}}{E_c-E_q} \Bigg|^{t_{g}}_0}_{=0} + \int_0^{t_{g}} dt \ e^{i\Phi(t)} \frac{d}{dt} \frac{\dot \theta}{E_c-E_q} \label{eqn:ibp}\\
    &= \int_0^{t_{g}} dt \ e^{i\Phi(t)} \frac{d}{dt}\frac{\dot \theta}{E_c-E_q}.
\end{align}
The transformed non-adiabatic error term is now in-phase with our control correction, allowing cancellation. 
From Eq. (\ref{eqn:Null_cond}), we see this is achieved by the choice:
\begin{equation}
    \delta \omega(t) = - \frac{2}{\sin 2\theta(t)} \frac{d}{dt} \left(\frac{\dot{\theta}(t)}{E_c(t)-E_q(t)}\right) \ .
    \label{eqn:corr1}
\end{equation}
A closed form of Eq. (\ref{eqn:corr1}) in terms of $\Delta$, $\dot \Delta$ and $\ddot \Delta$ can be found in Appendix \ref{sec:Convergence}. While the correction pulse $\delta \omega(t)$ found here eliminates the leakage errors (i.e.~transitions between adiabatic eigenstates), the control Hamiltonian in Eq. (\ref{eqn:H_ad}) also yields an additional diagonal component in the adiabatic eigenstate basis. As we see below, the resulting small phase error is easily mitigated by modifying the base pulse's hold detuning value $\Delta_h$ so that the gate condition in Eq. (\ref{eqn:gate_cond}) (including the control-induced phase contribution) is satisfied. Since this recalibration is typically on the order of a few MHz, the correction pulse still performs very well. For high numerical precision, the correction can also be iterated with the readjusted $\Delta_h$, which converges within very few steps.

Note that the use of integration by parts in our method has a strong analogy to its use in boundary cancellation methods to mitigate non-adiabatic errors \cite{Wiebe2012njp, Rezakhani2010PRA, Lidar2009JMP}, and can also be employed for a variety of other problems \cite{Ribeiro2017PRX}. It is also interesting to note a partial similarity to so-called net-zero pulses used to mitigate leakage in conditional-phase gates \cite{Rol2019PRL, Negirneac2021PRL}. Such pulses also engineer an end-of-gate leakage amplitude cancellation by interfering amplitudes generated in the first and second half of a pulse. This interference is only realized at discrete operating points, requiring joint calibration of pulse amplitude and timing alongside the gate condition. In contrast, our Magnus-based corrections enforce the cancellation analytically for a continuous family of pulse shapes and gate durations, even if the pulse shapes have no intrinsic symmetry.  Crucially, our approach is also directly applicable to non-diagonal gates like iSWAP, and as we show below, can be extended to mitigate multiple leakage processes.  

\subsection{Numerical simulation of coupler leakage suppression}
\label{sec:corr1_Simulation}

While we will extend our above discussion to include additional qubit levels, we first show numerically that the strategy of Sec.~\ref{sec:leak_corr} can be highly effective in suppressing coupler leakage.  
We employ a base pulse profile $\Delta(t)$ constructed using a smooth polynomial ramp function $P(x)$:
\begin{gather}
    \label{eqn:Poly}
    P(x) = 6x^5 - 15x^4 + 10x^3 \\
    \Delta(t) = \begin{cases}
        \Delta_0 + \delta\Delta \cdot P(t/\tau) & 0 \leq t < \tau \\
        \Delta_h & \tau \leq t \leq t_{g} - \tau \\
        \Delta_0 + \delta\Delta \cdot P((t_{g}-t)/\tau) & t_{g} - \tau < t \leq t_{g}
    \end{cases}
\end{gather}
where $\delta \Delta = \Delta_h - \Delta_0$ and $\tau$ denotes the ramp time;  as before, $\Delta_h$ ($\Delta_0$) is the hold (off-position) coupler detuning. 
 Values of $\Delta_0,g$ and $g_{12}$ are taken to match standard superconducting circuit parameters, and are given in Table \ref{tab:par_iSWAP}; $\Delta_h$ is varied (with $\tau,t_g$) to fulfill Eq. (\ref{eqn:gate_cond}) (thus ensuring a perfect iSWAP in the adiabatic limit). 

\begin{figure}[!tp]
    \centering
    \includegraphics[width = \columnwidth]{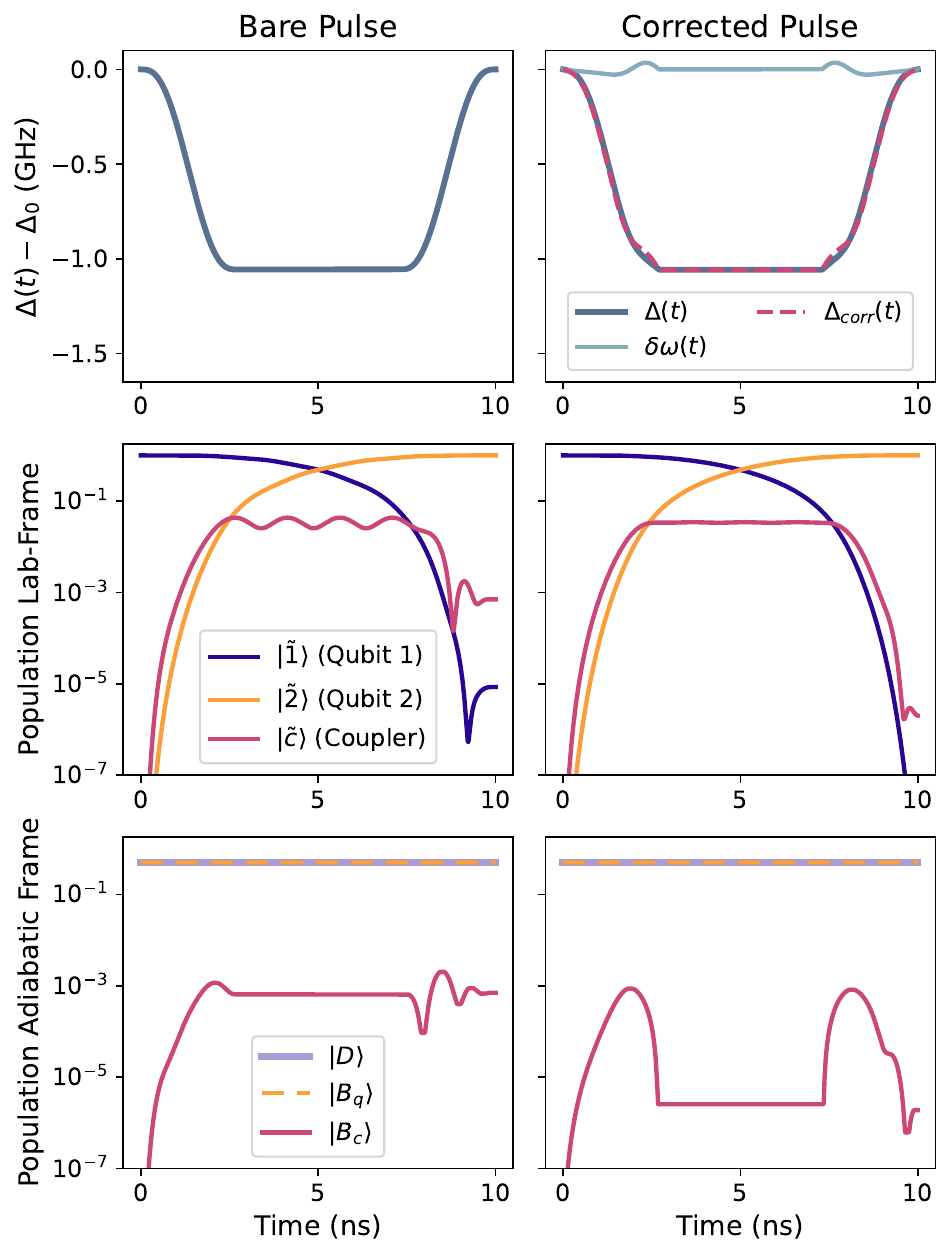}
    \caption{Simulation of an iSWAP gate in the one-excitation subspace (cf.~Eq. (\ref{eqn:H_lab})), comparing performance of a simple base coupler pulse (left column) with a corrected version (as derived in Eq. (\ref{eqn:corr1})) (right column). The corrected pulse effectively suppresses the interaction matrix element between the $\ket{B_q}$ and $\ket{B_c}$ states, reducing the population of the coupler-like state $\ket{B_c}$ in the adiabatic frame by several orders of magnitude (bottom row). Consequently, leakage into the coupler state at the end of the gate is significantly mitigated, as shown in the middle row. The simulation parameters are listed in Table \ref{tab:par_iSWAP}.}
    \label{fig:corr1}
\end{figure}

In Fig.~\ref{fig:corr1}, we simulate a rapid 10\,ns iSWAP gate based on the above pulse shape and using a ramp time $\tau = 2.7$\,ns. The left column shows performance using the bare detuning pulse, whereas the right column shows performance when the detuning pulse is modified by the additive correction $\delta \omega(t)$ of Eq.~(\ref{eqn:corr1}).  The top row shows both the bare pulse, and the corrected version. For the lower panels we simulate the time evolution of the computational state $\ket{\tilde 1}$, projected onto the lab-basis (middle row) and onto the adiabatic basis (bottom row). As in this simplified single-excitation setting the only error is leakage to the coupler, we consider an initial condition where at $t=0$ we are in the computational state $\ket{\tilde 1}$, and quantify populations in the final state. We see that by using our pulse correction, the residual leakage into the coupler is reduced by more than two orders of magnitude, i.e.~the final-state coupler population is $7.2\times 10^{-4}$ when using the bare pulse, but drops to $2.0\times 10^{-6}$ using the correction. For symmetry reasons, the results obtained when evolving the state $\ket{\tilde 2}$ are exactly the same. A full gate simulation of a realistic hardware Hamiltonian with multiple qubit levels can be found in Section \ref{sec:Hardware_result}. 

Fig.~\ref{fig:corr1} also provides intuition into how the correction operates: as shown in the bottom row, the pulse correction suppresses compared to the bare protocol the population of the coupler-like adiabatic bright state $\ket {B_c}$ both at the end of the protocol, but also during the middle ``hold" period of the protocol.  
The fact that there is still some residual leakage in the protocol stems from the perturbative nature of our correction, i.e.~due to our truncating the Magnus expansion to first order. 
\begin{table}[htpb]
    \centering
    \setlength{\tabcolsep}{2.0em} 
    \renewcommand{\arraystretch}{1.25} 
    \begin{tabular}{cc}
        \toprule
        \textbf{Parameters} &  \\
        \midrule
        $\alpha_1/2\pi = \alpha_2/2\pi = \alpha/2\pi$ & $-270\,\mathrm{MHz}$ \\
        $\alpha_c/2\pi$ & $-290\,\mathrm{MHz}$ \\
        $\Delta_0/2\pi$ & $1.522\,\mathrm{GHz}$ \\
        $g/2\pi$ & $180\,\mathrm{MHz}$ \\
        $g_{12}/2\pi$ & $21\,\mathrm{MHz}$ \\
        \bottomrule
    \end{tabular}
    \caption{System parameters used for the numerical simulation of Eq. (\ref{eqn:H_lab}) in Figs. \ref{fig:corr1}, \ref{fig:time_vs_fid} and \ref{fig:corr_sad} and Eq. (\ref{eqn:Hardware_H}) in Figs. \ref{fig:corr2},  \ref{fig:slepian} and \ref{fig:robustness}. The simulation outcomes are independent of the explicit value of the qubit frequency $\omega$. Note that $\Delta_0$ is not an independent parameter, but directly follows from Eq. (\ref{eqn:Delta_0}).}
    \label{tab:par_iSWAP}
\end{table}

\begin{figure}[!tp]
    \centering
    \includegraphics[width=\columnwidth]{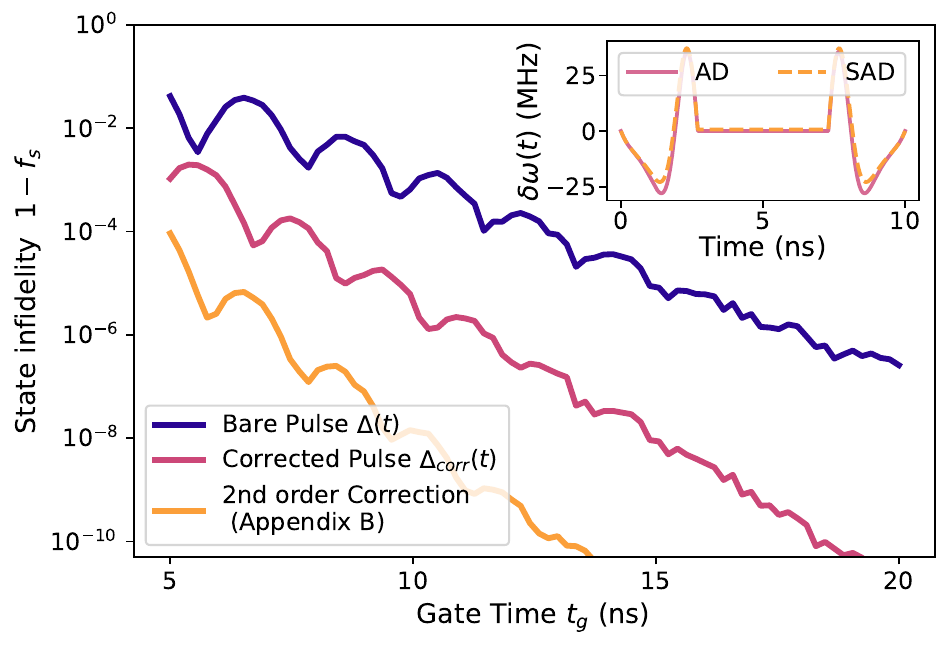}
    \caption{State infidelity $1-f_s = 1-|\braket{\psi_{\text{ideal}}|\psi(t_g)}|^2$ for an iSWAP gate simulated with the parameters of Table \ref{tab:par_iSWAP}. To account for realistic fabrication imperfections, the plotted results are averaged over normally distributed, correlated fluctuations of the coupling strengths (see main text for a detailed description). Over a wide range of gate times $t_g$, the first-order correction (pink line) reduces the state infidelity by at least two orders of magnitude compared to the bare pulse (blue line). Applying a higher-order correction in the superadiabatic frame (orange line, see Appendix \ref{sec:SAD_correction}) suppresses the error by approximately three more orders of magnitude, demonstrating the robustness and effectiveness of our correction scheme. Inset: Comparison of the correction $\delta \omega (t)$ derived in the adiabatic frame (AD) and the superadiabatic frame (SAD) for a 10\,ns gate.}
    \label{fig:time_vs_fid}
\end{figure}

It is also interesting to explore the efficacy of our correction as one varies the gate time $t_g$.  Fig.~\ref{fig:time_vs_fid} again shows results for a gate sequence starting from $\ket{\tilde 1}$ based on the above pulse shape and using the same parameters as in Table \ref{tab:par_iSWAP}, where we now vary $t_g$ and fix the relative ramp time $\tau / t_g = 0.27$ to match Fig.~\ref{fig:corr1}. We plot the state-infidelity $1-f_s = 1-|\braket{\psi_{\text{ideal}}|\psi(t_g)}|^2$, which is dominated by leakage errors.    
Further, to avoid fine-tuned features associated with accidental interference, for each $t_g$, we average over an ensemble of system parameters. The coupling strength is sampled from a Gaussian distribution as $g \sim \mathcal{N}(\mu_g, \sigma^2)$ with standard deviation $\sigma = 5\text{\,MHz}$ and mean value $\mu_g$ taken from Table \ref{tab:par_iSWAP}. The higher-order coupling is scaled with the identical relative deviation, $g_{12} = \mu_{g_{12}} \, g/\mu_g$, using the corresponding mean $\mu_{g_{12}}$ from Table \ref{tab:par_iSWAP}. The off-point $\Delta_0$ is calculated for each realization independently using Eq. (\ref{eqn:Delta_0}). 

Fig. \ref{fig:time_vs_fid} demonstrates that our  correction $\Delta_{\text{corr}}(t)$ (pink line) decreases the ensemble-averaged state infidelity by more than two orders of magnitude compared to the bare pulse $\Delta(t)$ (blue line) across the entire range of simulated gate times.  One can do even better by using a higher-order version of our correction. Interestingly, an effective strategy is to suppress the error in a higher-order adiabatic (superadiabatic) frame while remaining at first order in the Magnus expansion. A derivation of the corresponding pulse shape, along with a discussion of the error scaling, is provided in Appendix~\ref{sec:SAD_correction}. 
The resulting infidelity from this higher-order suppression is plotted in orange, and shows an additional improvement of three orders of magnitude.  

%% file: 3_Leakage_correction_in2exc.tex
\section{Leakage Correction in the two-excitation subspace}
\label{sec:leak_2subspace}
The previous section gave a minimal example of how a DRAG-style leakage correction could be derived for gates employing a single baseband control.  We now show how a similar approach can work in more complex settings.  We consider the same basic setup as Sec.~\ref{sec:LeakageSingleMain}, i.e.~a tunable coupler realizing an iSWAP gate.  We now however consider leakage errors in the two-excitation subspace, which is of course extremely relevant to realistic experiments.  The problem here is more complex because of multiple relevant leakage states.  Nonetheless, our approach is still feasible: we are able to derive simple correction pulses that mitigate leakage in this subspace. Further, in Sec. \ref{sec:two_channel_corr}, we will show that these can be augmented to find pulse corrections that simultaneously address leakage in both the one and two-excitation subspaces.   

The basic strategy is the same as in Sec.~\ref{sec:LeakageSingleMain}: we want a simple modulation to correct our detuning pulse $\Delta(t)$ so as to cancel leakage processes {\it on average}.  We will do this by applying the same three steps used in \ref{sec:leak_corr}: we first identify an effective low-dimensional Hamiltonian 
(cf.~Eq.~(\ref{eqn:2excH})), then transform into an interaction picture with respect to the ideal gate dynamics, and finally read off the correction from the requirement that the final-time leakage matrix element (obtained from a first-order Magnus expansion) vanishes.  

\subsection{Two-excitation leakage subspace}

The two-excitation manifold in principle contains six states, one of which is the computational state $\ket{101}$.  We have leakage states that involve the coupler (e.g.~$\ket{110}$), as well as leakage states only involving qubit excitations (e.g.~$\ket{200}$).  We start by making the approximation that in addition to having the same coupling $g$ to the coupler, both qubits also have the same anharmonicity parameter, i.e.~$\alpha_1 = \alpha_2 = \alpha$. A discussion of our pulse corrections in presence of parameter asymmetries can be found in Appendix \ref{sec:robustness}. We also ignore the state $\ket{020}$, assuming that the coupler detuning is always 
much larger than $\alpha$, making this a less-dominant leakage state.  Our leakage problem reduces to a three level subspace, as the computational state $\ket{101}$ can only couple to the symmetric leakage states
\begin{align}
    \label{eq:Symm2ExcitationStates}
    \ket{C_+}  &= \frac{1}{\sqrt{2}}(\ket{110} + \ket{011}) \notag \\
    \ket{Q_+}  &= \frac{1}{\sqrt{2}}(\ket{200} + \ket{002}) \ .
\end{align}
Using a basis $\{\ket{101}, \ket{C_+}, \ket{Q_+}\}$, the effective Hamiltonian for the two excitation subspace is:
\begin{equation}
\label{eqn:2excH}
    H_2 = \begin{pmatrix} 
        0 & \sqrt{2} g & 2 g_{12} \\ 
        \sqrt{2} g & \Delta(t) & \sqrt{2} g \\ 
        2 g_{12} & \sqrt{2} g & \alpha 
    \end{pmatrix} \ .
\end{equation}
We consider the same basic iSWAP gate protocol as in Sec.~\ref{sec:LeakageSingleMain}, where $\Delta(t)$ is ramped from the off-position $\Delta_0$ to a hold position $\Delta_h$.
Note that when the gate is off ($\Delta = \Delta_0$), there will be small residual $ZZ$ interaction due to the shift in the energy of the $\ket{101}$ state, as is standard for this style of gate, see Appendix \ref{sec:coupler_off} for further discussion.  While the gate will never involve strong hybridization of states in the two-excitation manifold, there will be a weak time-dependent dressing of states, and this time-variation can again cause non-adiabatic errors. 
We can again consider the instantaneous eigenstates of $H_2$, denoted using a tilde.  In the gate-off configuration, these eigenstates correspond to static dressed versions of our bare states.  We want to understand transitions between the adiabatic eigenstates during the gate, and pulse corrections that could mitigate this.  

\subsection{Approximate transformation to the adiabatic frame}

The first step in our strategy is to find the time-dependent adiabatic eigenstates, i.e.~move to the adiabatic frame.  Unfortunately, even with our $3\times3$ structure, exact expressions here become unwieldy. 
We will thus start by assuming $g$ is small enough to be treatable perturbatively in the two-excitation subspace, letting us approximately find the instantaneous eigenstates of $H_2$ via a standard Schrieffer--Wolff transformation.  This approximation will yield valuable physical insights, and sets the stage for Sec. \ref{sec:Hardware_result}, where we do not assume $g$ is perturbatively small, but instead implement our strategy in a  numerically-exact fashion.

Following the steps described in Sec. \ref{sec:leak_corr} we first transform the total system into the adiabatic frame associated with the bare pulse $\Delta(t)$, now just to leading order in $g$. This will let us identify the relevant non-adiabatic processes, and attempt a cancellation via a detuning modification $\Delta(t) \rightarrow \Delta(t) + \delta \omega(t)$.  
The total system Hamiltonian $H_2' = H_2 + H_{2,\text{ctrl}}$ is therefore transformed via a Schrieffer--Wolff transformation using the unitary $\hat U = \exp(\hat S)$ with its anti-Hermitian generator $\hat S$, defined as
\begin{equation}
\label{eqn:S_generator}
    S = \begin{pmatrix} 
        0 & \theta_1 & \theta_3 \\ 
        -\theta_1 & 0 & -\theta_2 \\ 
        -\theta_3 & \theta_2 & 0 
    \end{pmatrix} \ ,
\end{equation}
using the same basis as introduced above Eq. (\ref{eqn:2excH}). The dimensionless expansion parameters are given by
\begin{equation}
    \theta_1 = \frac{\sqrt{2}g}{\Delta(t)}, \,\,
    \theta_2 = \frac{\sqrt{2}g}{\Delta(t) - \alpha}, \,\,
    \theta_3 = \frac{2g_{12}}{\alpha} - \frac{\sqrt{2}g(\theta_1 + \theta_2)}{2\alpha}
\end{equation} 
Note that in typical transmon parameter regimes $g/\alpha \sim \mathcal{O}(1)$, and hence $\theta_3$ is the same order as $\theta_1$ and $\theta_2$. Expanding the exponential in powers of $S$ yields the transformed Hamiltonian as a series of commutators
\begin{widetext}
\begin{align}
    \hat H_{\text{ad}}' &= \hat U^\dagger \hat H' \hat U - i \hat U^\dagger \hat {\dot{U}} \\
    &\simeq H' + [H', S] + \frac{1}{2}[[H', S], S] - i\dot{S} + \frac{i}{2} [S, \dot{S}] + \mathcal{O}(\theta^3) \\
    &= \begin{pmatrix} 
        E_1(t) + \delta\omega \theta_1^2 & -(\delta\omega \theta_1 + i\dot{\theta}_1) & \delta\omega \theta_1\theta_2 + \frac{i}{2}(\dot{\theta}_1\theta_2 - \theta_1\dot{\theta}_2) - i\dot \theta_3 \\ 
        -(\delta\omega \theta_1 - i\dot{\theta}_1) & E_2(t) + \delta\omega(1 - \theta_1^2 - \theta_2^2) & -(\delta\omega \theta_2 - i\dot{\theta}_2) \\ 
        \delta\omega \theta_1\theta_2 - \frac{i}{2}(\dot{\theta}_1\theta_2 - \theta_1\dot{\theta}_2) + i\dot \theta_3 & -(\delta\omega \theta_2 + i\dot{\theta}_2) & E_3(t) + \delta\omega \theta_2^2 
    \end{pmatrix} \ ,
\end{align}
\end{widetext}
where $E_1(t) = -\sqrt{2}g\theta_1$, $E_2(t) = \Delta(t) + \sqrt{2}g(\theta_1 + \theta_2)$, and $E_3(t) = \alpha - \sqrt{2}g\theta_2$ are energies of the adiabatic eigenstates, perturbatively shifted due to hybridization.  As expected, we have terms coupling the adiabatic eigenstates, that stem from both, the non-adiabatic errors (terms involving $\dot{\theta}_j$), and the control correction $\delta \omega(t)$.  
There are in principle two relevant leakage channels involving the dressed computational state $\ket{\tilde{101}}$ (one to the qubit-like state $\ket{\tilde{Q}_+}$, one to the coupler-like state $\ket{\tilde{C}_+}$).  Motivated by the usual experimental situation where $|\alpha| < |\Delta(t)|$, we will focus exclusively here on cancelling leakage into the qubit-like state.    

To this end, we transform $H_{\text{ad}}'$ into the interaction picture with respect to the diagonal Hamiltonian, $H_0(t) = \text{diag}(E_1(t), E_2(t), E_3(t))$. Applying $U_I(t) = \exp\left(-i \int_0^t H_0(t_1) \, dt_1\right)$ yields $H_{\text{ad},I}'(t) = U_I^\dagger(t) H'_{\text{ad}}(t)U_I(t) - H_0(t)$, explicitly given by
\begin{widetext}
\begin{equation}
    H_{\text{ad},I}' = \begin{pmatrix} 
        \delta\omega \theta_1^2 & -(\delta\omega \theta_1 + i\dot{\theta}_1)e^{i\Phi_{12}(t)} & (\delta\omega \theta_1\theta_2 + \frac{i}{2}(\dot{\theta}_1\theta_2 - \theta_1\dot{\theta}_2) - i\dot \theta_3)e^{i\Phi_{13}(t)} \\ 
        -(\delta\omega \theta_1 - i\dot{\theta}_1)e^{-i\Phi_{12}(t)} & \delta\omega(1 - \theta_1^2 - \theta_2^2) & -(\delta\omega \theta_2 - i\dot{\theta}_2)e^{i\Phi_{23}(t)} \\ 
        (\delta\omega \theta_1\theta_2 - \frac{i}{2}(\dot{\theta}_1\theta_2 - \theta_1\dot{\theta}_2) + i\dot \theta_3)e^{-i\Phi_{13}(t)} & -(\delta\omega \theta_2 + i\dot{\theta}_2)e^{-i\Phi_{23}(t)} & \delta\omega \theta_2^2 
    \end{pmatrix} \ ,
\label{eqn:H_ad2exc}
\end{equation}
\end{widetext}
with the accumulated relative dynamical phases $\Phi_{jk}(t) = \int_0^t (E_j(t_1) - E_k(t_1)) \, dt_1$. 

\subsection{Correcting leakage to qubit-only states}

We are now in a position to formulate the conditions determining the control correction $\delta \omega(t)$.  As already stated, we will focus on just trying to mitigate the dominant leakage stemming from transitions between $\ket{101}$ and $\ket{Q_+}$.  Treating the transformed Hamiltonian via a first order Magnus-expansion, and then insisting that the resulting time-averaged Hamiltonian does not couple these two states, we obtain the condition: 
\begin{align}
\label{eqn:Null_cond2}
    0 \overset{!}{=} \int_0^{t_g} &\left( - \frac{i}{2}(\dot{\theta}_1\theta_2 - \theta_1\dot{\theta}_2) + i\dot{\theta}_3 + \theta_1\theta_2\delta\omega (t)\right) \notag \\
    &\times e^{-i\Phi_{13}(t)} \, dt \ .
\end{align}
Note that this condition targets the direct coupling element only. The indirect channel $\ket{\tilde{101}} \to \ket{\tilde{C}_+} \to \ket{\tilde{Q}_+}$ remains as a higher-order error of the same order as the neglected second Magnus term $\hat{\Omega}_2$.

We now need to solve this equation for $\delta \omega(t)$.  
Note the formal similarity to the condition determining the correction in the one-excitation subspace, Eq.~(\ref{eqn:Null_cond}).  Naively, we have the same problem: the $\delta \omega(t)$ term is $\pi/2$ out of phase with the other non-adiabatic error terms.  The solution is also similar: as we only need the integral to vanish (and not the integrand at each instant of time), we can find a real-valued solution for $\delta \omega(t)$ by integrating by parts.  Assuming a smooth $\Delta(t)$, whose derivatives vanish at the start and end of the gate, we can ignore the boundary terms. The resulting integral can be made zero via the choice:
\begin{equation}
    \delta\omega(t) = \frac{1}{2\theta_1\theta_2} \frac{d}{dt} \left( \frac{ -(\dot{\theta}_1\theta_2 - \theta_1\dot{\theta}_2) + 2\dot{\theta}_3 }{E_3(t) - E_1(t)} \right).
    \label{eqn:corr2}
\end{equation}
We have thus determined an approximate correction to our detuning pulse $\Delta(t)$ that corrects leakage to two-excitation qubit states to leading order.  Note that for standard transmon parameters 
(where $|\alpha| < |\Delta(t)|$), this correction will be parametrically larger than what was needed to correct leakage in the one-excitation subspace, since the dominant term for small $\alpha$ is the term $\propto \dot{\theta_3} \sim g/\alpha \cdot (\dot\theta_1 + \dot\theta_2)$.

While our perturbative treatment here helps elucidate the general procedure, for typical transmon setups, $g$ is not small enough to be treated perturbatively.  This motivates the next section, where we implement our strategy numerically without a small-$g$ assumption. 

%% file: 4_Results.tex
\section{Leakage correction in a realistic hardware setup}
\label{sec:Hardware_result}
\begin{figure}[!htp]
    \centering
    \includegraphics[width = \columnwidth]{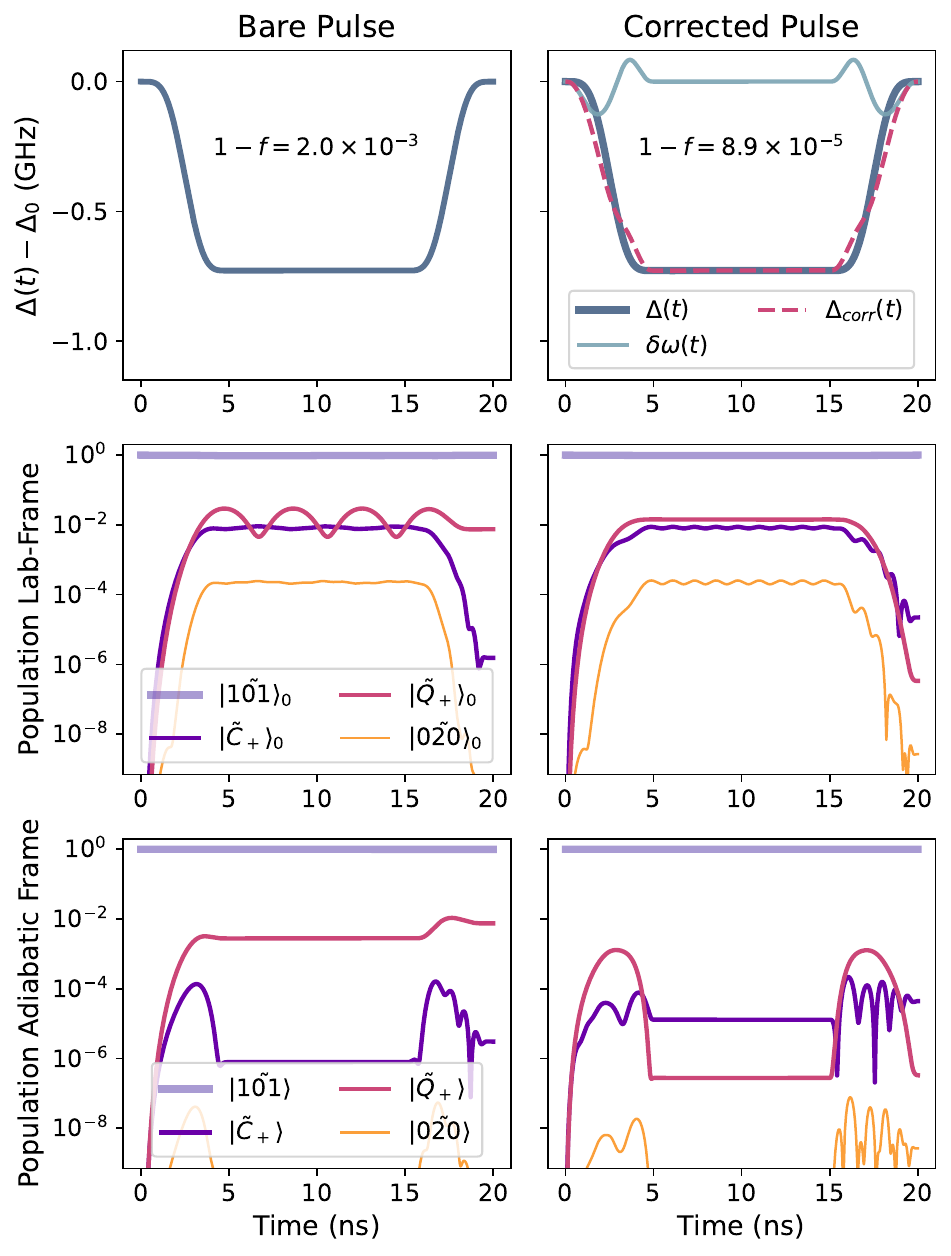}
    \caption{Simulation of Eq. (\ref{eqn:Hardware_H}) for an iSWAP gate, comparing a bare coupler pulse (left column) with the active, numerically precise correction derived in Section \ref{sec:leak_2subspace} (right column). Projected onto the computational basis, which corresponds to the eigenstates of Eq. (\ref{eqn:Hardware_H}) at $t=0$ (middle row), the population of the fidelity-limiting state $\ket{\tilde Q_+}_0$ drops from $\mathcal{O}(10^{-2})$ for the bare pulse to $\mathcal{O}(10^{-6})$ with the correction. This improvement stems from the suppression of the interaction matrix element in the adiabatic frame, which reduces the population of the time-dependent instantaneous eigenstate $\ket{\tilde Q_+}$ by three orders of magnitude (bottom row). Both the bare and corrected pulses are plotted in the top row. The simulation parameters are listed in Table \ref{tab:par_iSWAP}.}
    \label{fig:corr2}
\end{figure}

Secs.~\ref{sec:LeakageSingleMain} and \ref{sec:leak_2subspace} illustrated the principle and efficacy of our general leakage mitigation strategy in somewhat simplified settings.  We now apply our correction strategy to a more complete model of a flux-tuned two-qubit gate, showing that our approach remains feasible and effective.  
As is standard, we model each transmon qubit and the coupler itself as anharmonic oscillators (anharmonicity parameters $\alpha_j$).  We also drop excitation non-conserving terms corresponding to a rotating wave approximation, since they are not resonantly driven by the baseband flux pulse and merely renormalize the system parameters. This yields the full system Hamiltonian:
\begin{align}
    \hat H &= \sum_{i \in \{1, c, 2\}} \left( \omega_i \hat a_i^\dagger \hat a_i + \frac{\alpha_i}{2} \hat a_i^\dagger \hat a_i^\dagger \hat a_i \hat a_i \right) \notag \\
    &+ \sum_{i \in \{1,2\}} g \left(\hat a_i^\dagger \hat a_c + \text{H.c.}\right)  + g_{12} \left(\hat a_1^\dagger \hat a_2 + \text{H.c.}\right) \ .
    \label{eqn:Hardware_H}
\end{align}
We take both qubits to be resonant and work in a frame rotating at the qubit frequency, hence $\omega_1 = \omega_2 = 0$ and $\omega_c = \Delta(t)$. In what follows, we retain only three energy levels for each subsystem, as this captures the dominant coherent error pathways. Note that the use of a truncated nonlinear oscillator model is not necessary: our approach could work equally well for more realistic models (and can be extended to include additional levels).    

As before, the gate is realized via a time-dependent coupler--qubit detuning that has the form of a bare pulse $\Delta(t)$ plus a small correction $\delta\omega(t)$ designed to mitigate leakage. For our simulations we use $\Delta (t)$ defined in Eq. (\ref{eqn:Poly}). While there are in principle many leakage channels, for realistic parameters the dominant channel involves the two-excitation qubit state $\ket{\tilde Q_+}_0$. This state is defined as the dressed eigenstate of the Hamiltonian in Eq. (\ref{eqn:Hardware_H}) at $t=0$ that corresponds to the bare state $\ket{Q_+}$ (cf.~Eq.~(\ref{eq:Symm2ExcitationStates})).  We will thus start by designing a correction to mitigate this leakage channel.

Our correction strategy will again be based on moving to the adiabatic frame, and canceling the relevant non-adiabatic leakage error on average with a pulse modification.  While the strategy is the same as in Sec.~\ref{sec:leak_2subspace}, we will now use exact numerical expressions for the instantaneous eigenstates, hence going beyond the approximate perturbative form in Eq.~(\ref{eqn:corr2}).
We stress that the numerics required here reduce to diagonalization of low-dimensional matrices, and do not require anything close to the effort required for full numerical optimal control.  See Appendix \ref{sec:numerics} for more details on the explicit derivation of the correction.  

To evaluate the impact of coherent errors on gate performance and directly assess the utility of our pulse correction, we calculate the gate fidelity within the logical subspace \cite{Pedersen2007PLA, Kueng2016PRL}, defined as
\begin{equation}
\label{eqn:fidelity}
    f = \frac{ |\text{Tr} (\tilde U^\dagger U_{\text{ideal}})|^2 + |\text{Tr} (\tilde U^\dagger \tilde U)|}{d \cdot (d+1)}
\end{equation}
where $U_{\text{ideal}}$ is the ideal iSWAP unitary and $d=4$ is the dimension of the computational subspace. The matrix $\tilde U$ is the projection of the full propagator on the four computational states $\ket{\tilde{000}}_0$, $\ket{\tilde{100}}_0$, $\ket{\tilde{001}}_0$, and $\ket{\tilde{101}}_0$ with the entries defined as $\tilde U_{ij} = \bra{i} \hat{\mathcal{T}} \exp (-i\int_0^{t_g} dt' H(t')) \ket{j}$ and the evolution generated by the time-dependent Hamiltonian in Eq. (\ref{eqn:Hardware_H}).
As is standard, we also optimize over final single-qubit $Z$ gates to cancel trivial single-qubit phase errors (something that can be implemented easily via virtual frame updates \cite{McKay2017PRA}). 
Analogous to the procedure in 
Sec.~\ref{sec:corr1_Simulation}, for a given gate time $t_g$, we optimize the hold-value of the detuning $\Delta_h$ to maximize gate fidelity; this is done independently for both the original bare detuning pulse $\Delta(t)$ and the corrected version $\Delta_{\text{corr}}(t)$.
We also use system parameters listed in Table \ref{tab:par_iSWAP}. 

Results for our simulations (with and without a correction) for a fast $20\text{\,ns}$ iSWAP gate are presented in Fig.~\ref{fig:corr2}.  These simulations include three levels each for the coupler and both qubits, and include all coupling terms in the Hamiltonian of Eq.~(\ref{eqn:Hardware_H}). 
We find that our correction improves the gate infidelity
$1-f$ by more than an order of magnitude, from $2.0\times10^{-3}$ for the bare pulse to $8.9\times10^{-5}$ when applying the correction. 
Note that this suggests our correction reduces coherent leakage errors to a level where they no longer matter, as they will be dominated by incoherent errors. For example, for typical transmon relaxation times $T_1 \sim 50\,\mu\mathrm{s}$, one would obtain an additional incoherent error $\sim 10^{-4}$ \cite{Sung2021PRX, Collodo2020PRL}.  
We stress that unlike leakage, standard qubit $T_1$ and $T_2$ processes yield errors within the computational subspace, and hence can be addressed by standard QEC codes (see e.g.~\cite{Aliferis2007QIC, Suchara2015QIC, Google2024Nature, Krinner2022Nature}).

Fig.~\ref{fig:corr2} demonstrates that without correction, the gate has significant final leakage to the qubit-like state $\ket{\tilde Q_+}_0$, and a smaller leakage to the coupler-like state $\ket{\tilde C_+}_0$.  Our correction targets the first process, dramatically reducing its value.  Gate infidelity is thus limited by the residual leakage to $\ket{\tilde C_+}_0$.  This is also something that can be addressed with a modified version of our approach: in the next section, we demonstrate an extended strategy that allows one to simultaneously target two distinct leakage channels.  As we show, this allows an even greater enhancement of gate fidelity.  

Note that the basic correction strategy presented here can be extended to various two-qubit gates activated by baseband pulses. An additional example of a CZ gate, based on the implementation in \cite{Collodo2020PRL}, is provided in Appendix \ref{sec:CZ_correction}. Furthermore, in Section \ref{sec:Slepian}, we benchmark our results against the widely used Slepian-based optimal control strategy \cite{Martinis2014PRA, Sung2021PRX}. We also demonstrate how our analytical methods can be utilized to overcome intrinsic limitations of these pulse shapes.

\section{Pulse corrections to mitigate multiple leakage pathways}
\label{sec:two_channel_corr}

The previous section highlights a general problem: in many settings there are multiple relevant leakage states, while our basic strategy mitigates a single pathway. We discuss here how to go beyond this limitation, again focusing on the two-qubit setup in Sec. \ref{sec:Hardware_result}. 

\subsection{Basic sequential strategy}

The most naive approach when faced with multiple leakage channels would be to use Eq. (\ref{eqn:Null_cond}) to derive corrections for each channel, and then simply sum these to get a total pulse correction.  This approach is highly non-optimal, as the different corrections will disrupt one another.  
We instead take an alternate, sequential approach, using the fact that in many problems, there is a hierarchy in the magnitude of different leakage processes.  We first derive a correction $\delta \omega^{(1)}(t)$ designed to {\it just} mitigate the dominant leakage pathway.  Then, we derive an additional correction $\delta \omega^{(2)}(t)$ to mitigate a second leakage channel, now including the effects of $\delta \omega^{(1)}(t)$, and in a way that ensures $\delta \omega^{(2)}(t) \ll \delta \omega^{(1)}(t)$.  This simple strategy helps minimize unwanted competition between corrections and (as we show) can be highly effective.    

We implement this general idea for the same two-qubit gate scenario as Sec. \ref{sec:Hardware_result}. We first find a correction $\delta \omega^{(1)}(t)$ that mitigates leakage to $\ket{\tilde Q_+}$
(cf.~Eq.~(\ref{eq:Symm2ExcitationStates})), using the approach described in Sec. \ref{sec:leak_2subspace}. We next want to find an additional pulse correction $\delta \omega^{(2)}(t)$ that suppresses the remaining leakage to $\ket{\tilde C_+}_0$.  This must be done under the constraint $|\delta \omega^{(2)}(t)| \ll |\delta \omega^{(1)}(t)|$, so that this second correction does not disrupt the first.  The cancellation condition determining $\delta \omega^{(2)}(t)$ is obtained from the dressed Hamiltonian in Eq. (\ref{eqn:H_ad2exc}), and is 
analogous to Eq. (\ref{eqn:Null_cond2}):
\begin{equation}
    0 \overset{!}{=} \int_0^{t_{g}} \left( i\dot{\theta}_1 - \theta_1 \bigl( \delta\omega^{(1)} + \delta\omega^{(2)} \bigr) \right) e^{-i\Phi_{12}(t)} \, dt \ .
\end{equation}

We could solve this using a single integration by parts, as we did for $\delta \omega^{(1)}$.  However, such an approach would not satisfy our constraint 
$|\delta \omega^{(2)}(t)| \ll |\delta \omega^{(1)}(t)|$. To enforce this hierarchy we instead apply integration by parts to the interaction-picture integral {\it multiple} times.  This results in the equation:
\begin{align}
\delta E (t) &= E_2(t)-E_1(t) \sim \Delta (t) \notag \\
    \delta \omega^{(2)} &= \frac{1}{\theta_1}\frac{d}{dt} \Biggl(\frac{1}{\delta E (t)} \notag \\
    &\times \frac{d}{dt}\Biggl( \frac{1}{\delta E (t)} \left( \theta_1 \delta \omega^{(1)} + \frac{d}{dt}\frac{\dot\theta_1}{\delta E (t)}\right)\Biggr)\Biggr) \ ,
\label{eqn:delta_omega_2}
\end{align}
where $E_1(t), E_2(t)$ are defined in Sec. \ref{sec:leak_2subspace}. Each additional integration-by-parts yields an additional derivative of the base pulse $\Delta(t)$, while at the same time generates an additional factor of $1/\Delta(t)$.  The result is a suppression of the magnitude of the pulse correction by a factor $\sim\dot\Delta/\Delta^2$ with each additional integration by parts. As each step also introduces a phase factor 
$i$, the real (imaginary) parts of the pulse correction need to be integrated two (three) times to ensure a final real-valued pulse.   

We stress that the approach here puts stronger smoothness constraints on the original base pulse $\Delta(t)$, as all boundary terms generated by the repeated partial integrations must vanish.  The result is that the base pulse now has to be at least $\mathcal{C}^4$-continuous. 
Using the above procedure and in direct analogy to  Sec.~\ref{sec:leak_2subspace}, we can obtain a numerically precise secondary correction $\delta\omega^{(2)}\ll\delta\omega^{(1)}$ that cancels the residual leakage into $|\tilde C_+\rangle_0$ without spoiling the nulling condition for $|\tilde Q_+\rangle_0$ to leading order (see Appendix~\ref{sec:numerics} for details).

\begin{figure}
    \centering
    \includegraphics[width=\columnwidth]{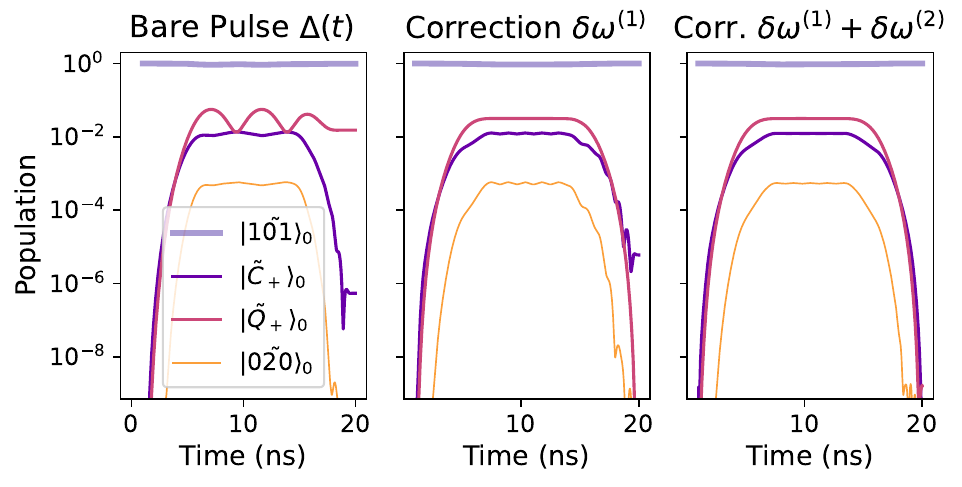}
    \caption{State populations in the second-excited subspace during the 20\,ns iSWAP gate. The dynamics of Eq. (\ref{eqn:Hardware_H}) was simulated using three levels per qubit for three distinct scenarios: using the bare coupler pulse alone (left), with the primary correction $\delta \omega^{(1)}$ (middle), and with the combined correction $\delta \omega^{(1)} + \delta \omega^{(2)}$ to suppress two leakage states (right). To optimally balance the suppression of the residual $\ket{\tilde C_+}_0$ population against higher-order errors induced by the secondary correction, the individual amplitudes of $\delta \omega^{(1)}$ and $\delta \omega^{(2)}$ were optimized as independent free parameters for the results shown in the right panel.}
    \label{fig:corr_2AD}
\end{figure}
\begin{table}[tpb]
    \centering
    \setlength{\tabcolsep}{2.0em} 
    \renewcommand{\arraystretch}{1.25} 
    \begin{tabular}{cc}
        \toprule
        \textbf{Parameters} &  \\
        \midrule
        $\alpha_1/2\pi = \alpha_2/2\pi = \alpha/2\pi$ & $-220\,\mathrm{MHz}$ \\
        $\alpha_c/2\pi$ & $-190\,\mathrm{MHz}$ \\
        $\Delta_0/2\pi$ & $1.414\,\mathrm{GHz}$ \\
        $g/2\pi$ & $165\,\mathrm{MHz}$ \\
        $g_{12}/2\pi$ & $19\,\mathrm{MHz}$ \\
        \bottomrule
    \end{tabular}
    \caption{System parameters used for the numerical simulation of Eq. (\ref{eqn:Hardware_H}) in Fig. \ref{fig:corr_2AD}. The simulation outcomes are independent of the explicit value of the qubit frequency $\omega$. In contrast to Table \ref{tab:par_iSWAP}, which was selected to give a well-performing iSWAP gate including residual $ZZ$ errors, this set was selected such that the hierarchically smaller second correction is clearly resolvable. Absolute error values in the two sections are therefore not directly comparable.}
    \label{tab:par_2AD}
\end{table}

\subsection{Application to a fast coupler-mediated gate}

\begin{table}[htbp]
    \centering
    \small
    \setlength{\tabcolsep}{6pt}
    \renewcommand{\arraystretch}{1.3}
    \begin{tabular}{cccc}
        \toprule
        Pulse & iSWAP $1-f$ & fSim $1-f$ & $L_1$ \\
        \midrule
        $\Delta(t)$ & $5.7\times10^{-3}$ & $3.9\times10^{-3}$ & $3.9\times10^{-3}$ \\
        
        $+\,\delta\omega^{(1)}$ & $1.5\times10^{-3}$ & $3.2\times10^{-6}$ & $3.2\times10^{-6}$ \\
        $+\,\delta\omega^{(1)}+\delta\omega^{(2)}$ & $1.5\times10^{-3}$ & $1.1\times10^{-9}$ & $1.0\times10^{-9}$ \\
        \bottomrule
    \end{tabular}
    \caption{Summary of gate infidelities $1-f$ and leakage rate $L_1$ \cite{Wood2018PRA} for the $20\,$ns iSWAP gate of Fig. \ref{fig:corr_2AD}, for the parameters of Table \ref{tab:par_2AD}. The corrections $\delta \omega^{(1)}$ ($\delta \omega^{(2)}$) are the numerically exact implementations of Eq. (\ref{eqn:corr2}) (Eq. (\ref{eqn:delta_omega_2})).}
    \label{tab:fidelities}
\end{table}

As an example of how this approach to multiple leakage pathways can be effective, we show in Fig. \ref{fig:corr_2AD} results of simulations of a fast $20\,\rm{ns}$ iSWAP gate via the Hamiltonian in 
Eq.~(\ref{eqn:Hardware_H}), using corrections that address two leakage states. The main results are summarized in Table \ref{tab:fidelities}. We discuss this in detail in what follows. 

To satisfy the requirement of vanishing derivatives up to the fourth order at the boundaries, we use an appropriately smooth base pulse. We replace the standard ramp polynomial used earlier (Eq.~(\ref{eqn:Poly})) with the following profile:
\begin{align}
\label{eqn:Poly2}
    P(x) = 70x^9 - 315x^8 + 540x^7 - 420x^6 + 126x^5 \ .
\end{align}
We first explore the improvement generated by using a pulse correction 
$\delta \omega^{(1)}(t)$ that only addresses the dominant leakage state (as derived in Sec.~\ref{sec:leak_2subspace}). In this case, however, we use the smoother base pulse defined in Eq.~(\ref{eqn:Poly2}) and the parameter set listed in Table~\ref{tab:par_2AD}.
As illustrated in the left panel of Fig. \ref{fig:corr_2AD}, applying the uncorrected base pulse leaves a substantial residual population in the $\ket{\tilde Q_+}_0$ state, limiting the overall gate infidelity to $5.7 \times 10^{-3}$. Implementing a correction $\delta \omega^{(1)}(t)$ to just address the dominant leakage state successfully suppresses this $\ket{\tilde Q_+}_0$ leakage, reducing the infidelity to $1.5 \times 10^{-3}$.

One might be surprised that fidelity is not improved further by our pulse correction $\delta \omega^{(1)}(t)$.  There is a simple reason for this: the dominant contribution to the remaining infidelity after pulse correction is no longer leakage, but a phase error due to a residual effective $ZZ$ interaction. 
This phase error could in principle be corrected using a standard QEC code.  This motivates including the residual $ZZ$ phase $\phi$ in the definition of our target gate, and hence computing the fidelity of our gate with an ideal fSim$(\pi/2, \phi)$ gate, given by
\begin{equation}
U_\text{fSim}(\theta, \phi) = \begin{pmatrix}
    1 & 0 & 0 & 0 \\
    0 & \cos \theta & -i\sin \theta & 0 \\
    0 & -i\sin \theta & \cos \theta & 0 \\ 
    0 & 0 & 0 & \exp(-i\phi)
\end{pmatrix} \ .
\label{eqn:fSim}
\end{equation}
This infidelity (which essentially removes the $ZZ$ phase error) is now significantly lower, $3.2 \times 10^{-6}$, and is almost entirely due to 
the second leakage pathway that we have not addressed: leakage to the coupler state $\ket{\tilde{C}_+}_0$. 

To more carefully quantify how much of this residual error is leakage rather than a coherent error within the computational subspace, we use the leakage rate \cite{Wood2018PRA}
\begin{equation}
    L_1 = 1 - \frac{\mathrm{Tr}(\tilde U^\dagger \tilde U)}{d} \ ,
    \label{eqn:leakage_rate}
\end{equation}
with $\tilde U$ (a non-unitary matrix defined below Eq. (\ref{eqn:fidelity})) and $d$ the same as in Eq.~(\ref{eqn:fidelity}).  This quantifies the population lost from the computational subspace, averaged over initial computational basis states. Applying the correction $\delta\omega^{(1)}(t)$ reduces the leakage rate from $L_1 = 3.9 \times 10^{-3}$ to $L_1 = 3.2 \times 10^{-6}$. Since $L_1 \leq 1-f$ holds in general, with equality only if the deviation from the target gate consists exclusively of population loss, the agreement between the two values shows that the phase-corrected gate is limited entirely by residual leakage. The bottleneck has shifted from residual population in $\ket{\tilde Q_+}_0$ to $\ket{\tilde C_+}_0$, at a level more than three orders of magnitude lower.

We now finally turn to our improved strategy that simultaneously mitigates two leakage pathways, i.e.~derive a second correction via Eq. (\ref{eqn:delta_omega_2}).
Here, this approach allows us to mitigate leakage both to the state $\ket{\tilde Q_+}_0$ and to the state $\ket{\tilde C_+}_0$. 
To balance the error from the second-order Magnus term in channel $\ket{\tilde C_+}_0$ against the crosstalk error $\delta \omega^{(2)}$ in the $\ket{\tilde Q_+}_0$ error channel, we multiply both corrections by scale factors, i.e., $A_1 \cdot \delta \omega^{(1)} + A_2 \cdot \delta \omega^{(2)}$, which are optimized by allowing for a slight deviation from unity. This minimizes the overall leakage error in this setting.

As seen in the right panel of Fig. \ref{fig:corr_2AD}, this dual-channel suppression removes the residual coupler population, ultimately yielding an improved gate infidelity of $1.1 \times 10^{-9}$ and the leakage rate drops down to $L_1 = 1.0\times 10^{-9}$. The remaining error is thus still leakage limited, now at a level far below typical decoherence thresholds. These results are summarized in Table \ref{tab:fidelities}.

%% file: 5_Slepian.tex
\section{Improvement of Slepian-based pulse shapes}
\label{sec:Slepian}
\begin{figure}[!htp]
    \centering
    \includegraphics[width=\columnwidth]{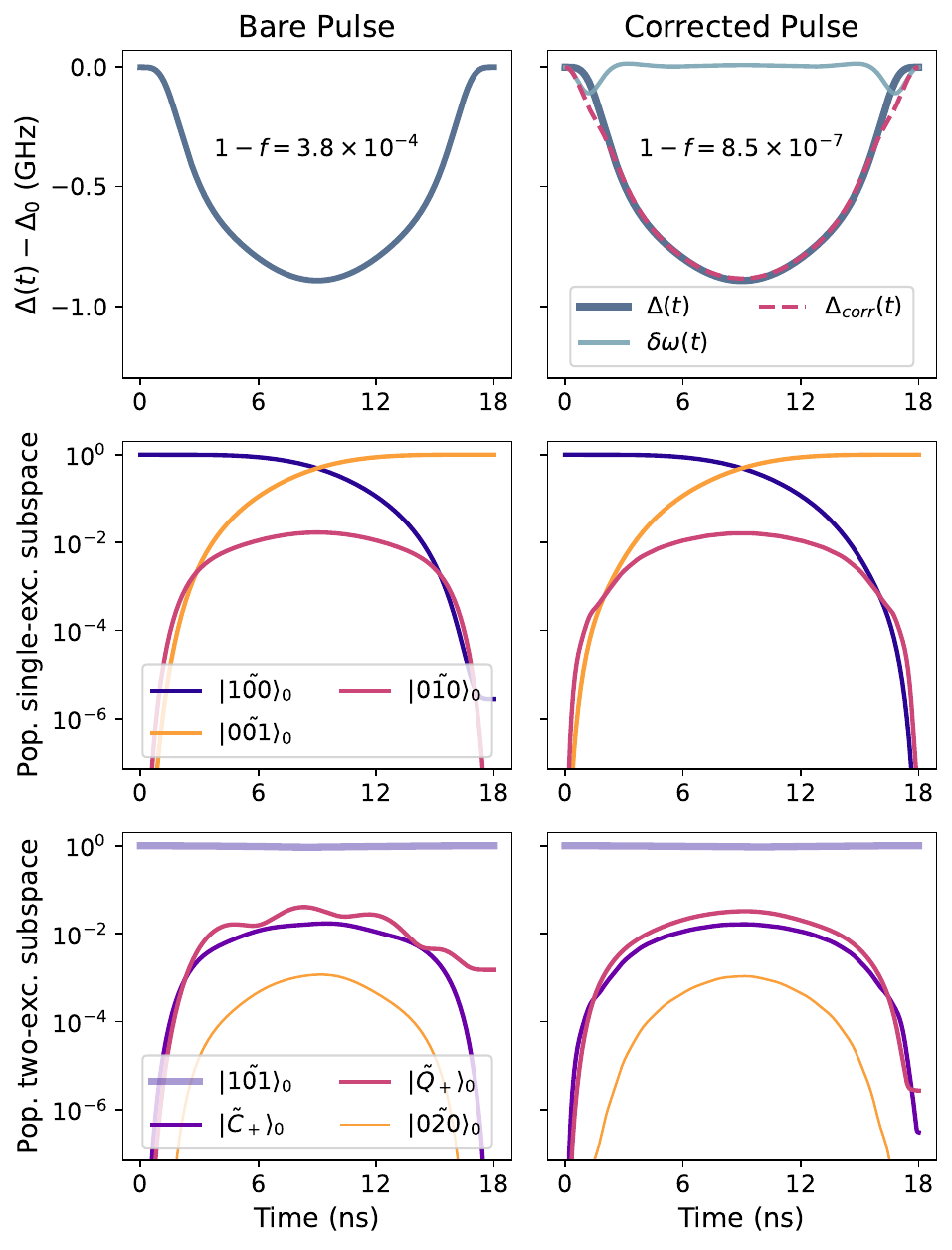}
    \caption{Simulation of the iSWAP gate dynamics using a bare Slepian-based pulse (left column) with our analytically corrected pulse (right column). The top panels display the respective coupler detunings $\Delta(t)$. In the single-excitation subspace (middle panels), both pulses successfully suppress residual leakage into the coupler state $\ket{\tilde{010}}_0$. In the two-excitation subspace (bottom panels), however, the bare Slepian pulse fails to suppress leakage into the strongly dressed $\ket{\tilde{Q}_+}_0$ state, leaving a residual population of $\sim 10^{-3}$. Applying our analytical correction selectively depopulates this specific leakage channel by several orders of magnitude while preserving the Slepian protection for all other states. Simulation parameters are provided in Table \ref{tab:par_iSWAP}; the infidelities are measured with respect to the $U_{\text{fSim}}$-gate, defined in Eq. (\ref{eqn:fSim}).}
    \label{fig:slepian}
\end{figure}

In this section, we demonstrate that our pulse shaping technique can easily be combined with established methods to mitigate gate errors. Since our approach is highly flexible regarding the underlying base pulse, any strategy for deriving such a suitable base pulse can be used and further improved by applying our modification without sacrificing the general benefits of the chosen pulse. As a concrete demonstration, we apply our approach to the optimal control strategy based on Slepian pulse shapes, which was proposed in \cite{Martinis2014PRA} and experimentally implemented in \cite{Sung2021PRX}. Their strategy relies on truncating the leakage subspace to an effective three-level system containing only the directly coupled states $\ket{110}$ and $\ket{011}$. This effective subspace is then rotated such that a bright state $\ket{C_+}$ couples as a single effective leakage state to the computational state, while a dark state $\ket{C_-}$ is decoupled and interacts merely via second-order processes \cite{Sung2021PRX}. 

Within this effective two-level system, leakage can be mitigated by a Slepian-based pulse \cite{Martinis2014PRA}, constructed in a fictitious time frame in which the transition frequency is constant and the spectral weight driving non-adiabatic transitions is minimized. We approximate this shape by a truncated Fourier cosine series with four coefficients and an additional $\mathcal C^2$ constraint, and map it back onto the physical detuning $\Delta (t)$ following \cite{Martinis2014PRA}.

Simulating Eq. (\ref{eqn:Hardware_H}) using the geometric Slepian strategy for our base pulse $\Delta (t)$ we find that this works exceptionally well for suppressing leakage into directly coupled coupler states, as demonstrated in the left column of Fig. \ref{fig:slepian}. The population of the coupler state $\ket{\tilde{010}}_0$ in the single-excitation subspace (middle panel), as well as the bright state $\ket{\tilde{C}_+}_0$ in the two-excitation subspace (lower panel), are strongly suppressed. 

The fundamental limitation of this strategy, however, lies in the initial truncation of the second-excited subspace, which strictly requires the condition $|g| \ll |\alpha|$ \cite{Sung2021PRX}. In typical transmon architectures, and specifically in our parameter regime, this condition is heavily violated, leading to a strong hybridization of the higher-order states. Consequently, the low-pass filtering of the Slepian pulse is insufficient to protect the strongly dressed state $\ket{\tilde{Q}_+}_0$. For the standard iSWAP gate, the population of this state, combined with residual $ZZ$-interactions, limits the gate infidelity to $5.4 \times 10^{-4}$ for an $18\,\rm{ns}$ gate. 
To properly isolate and evaluate the remaining leakage error without the interference of $ZZ$-errors, we analyze the $\text{fSim}(\pi/2, \phi)$-gate (as introduced in Sec. \ref{sec:two_channel_corr}). By optimizing the $ZZ$-angle $\phi$, we effectively remove the phase errors, revealing that the gate performance is ultimately limited by the leakage into $\ket{\tilde{Q}_+}_0$, resulting in an infidelity of $3.8 \times 10^{-4}$.

To overcome this limitation, we apply the analytical correction strategy described in Sec. \ref{sec:leak_2subspace} to actively target the remaining error in the $\ket{\tilde{Q}_+}_0$ state. As shown in the right column of Fig. \ref{fig:slepian}, our derived correction (Eq. (\ref{eqn:corr2})) for the underlying Slepian base pulse is capable of depopulating this leakage state by several orders of magnitude. Crucially, superimposing this correction has only a negligible impact on the Slepian pulse's ability to protect the coupler states. This combined approach yields a final gate infidelity of $8.5 \times 10^{-7}$ for the $\text{fSim}(\pi/2, \phi)$-gate, representing an improvement by nearly three orders of magnitude over the standalone Slepian strategy and pushing the theoretical leakage limits well below typical decoherence thresholds.

%% file: 6_Conclusion.tex
\section{Conclusions}
\label{sec:Conclusions}
In this work, we have demonstrated that the apparent control deficit in single-parameter quantum architectures is not a fundamental physical limitation. Rather than relying on additional hardware lines, our approach unlocks hidden control degrees of freedom by transforming the error dynamics via integration by parts. This mathematical reframing effectively rotates the otherwise uncorrectable diabatic perturbations onto the axis of our available control.

This strategy is a highly versatile and hardware-efficient solution to the pervasive problem of diabatic leakage. Without requiring complex hardware modifications, our analytically derived pulses can be flexibly adapted to selectively suppress a wide variety of leakage channels. Furthermore, the operational overhead remains minimal, demanding only $\mathcal{C}^2$-smoothness of the underlying base pulse. As validated by our numerical simulations, strategically leveraging the hierarchical dynamics of higher-order adiabatic frames suppresses residual leakage by several orders of magnitude. This translates directly into drastically improved two-qubit gate fidelities across broad parameter regimes, proving that severe architectural bottlenecks can be effectively overcome through physics-informed, analytical pulse engineering.

\vspace{1cm}

\begin{acknowledgments}
This work was supported by the Army Research Office under Grant No. W911NF-23-1-0077, and by the Simons Foundation through a Simons Investigator Award (Grant No.~669487).
This work also received funding from the German Federal Ministry of Education and Research via the funding program quantum technologies - from basic research to the market under contract number 13N16182 "MUNIQC-SC" and is part of the Munich Quantum Valley, which is supported by the Bavarian state government with funds from the Hightech Agenda Bayern Plus.

\end{acknowledgments}

%% file: Appendix.tex
\section{Decoupling point of the coupler}
\label{sec:coupler_off}
In this section we demonstrate how the condition $\Delta_0 = -g_{12} + g^2/g_{12}$ in Eq. (\ref{eqn:Delta_0}) leads to a decoupling of the qubit states. In contrast to standard procedures using perturbative Schrieffer--Wolff techniques \cite{Yan2018PRApplied}, this condition can be found using the exact eigenstates of the single-excitation subspace Hamiltonian in Eq. (\ref{eqn:H_lab}). Inserting this condition into Eq. (\ref{eqn:Eq}), the radicand collapses to a perfect square,
\begin{align}
    E_q(\Delta_0)
    &= \frac{1}{2}\left(\frac{g^2}{g_{12}}
       - \sqrt{\left(2g_{12}-\frac{g^2}{g_{12}}\right)^{\!2} + 8g^2}\,\right)
       \nonumber \\
    &= \frac{1}{2}\left(\frac{g^2}{g_{12}}
       - \sqrt{\left(2g_{12}+\frac{g^2}{g_{12}}\right)^{\!2}}\,\right)
       \nonumber \\
    &= -g_{12} = E_D \ .
\end{align}
The qubit-like bright state is therefore exactly degenerate with the dark state, and any basis of the two-dimensional eigenspace $\mathrm{span}\{\ket{B_q(\theta_0)}, \ket{D}\}$ is an equally valid eigenbasis. We use this freedom to pick the combinations in Eqs. (\ref{eqn:tilde1}) and (\ref{eqn:tilde2}), which are the ones localized on the individual qubits and thus carry no residual exchange interaction. We can explicitly calculate these states 
\begin{align}
|\tilde 1\rangle &= \frac{1}{\sqrt 2}\Big(\tfrac{\cos\theta_0}{\sqrt 2}\big(|1\rangle+|2\rangle\big)
 + \sin\theta_0\,|c\rangle + \tfrac{1}{\sqrt 2}\big(|1\rangle-|2\rangle\big)\Big) \notag \\
 &=\frac{1+\cos\theta_0}{2}\,|1\rangle + \frac{\sin\theta_0}{\sqrt 2}\,|c\rangle + \frac{\cos\theta_0-1}{2}\,|2\rangle \ , \\
|\tilde 2\rangle &= \frac{\cos\theta_0-1}{2}\,|1\rangle + \frac{\sin\theta_0}{\sqrt 2}\,|c\rangle + \frac{1+\cos\theta_0}{2}\,|2\rangle \ ,
\end{align}
with $\theta_0$ introduced in Sec. \ref{sec:single_exc_subspace}. 

The decoupling condition Eq.~(\ref{eqn:Delta_0}) removes the exchange interaction, but it does not by itself remove the residual $ZZ$ crosstalk $\zeta = \tilde E_{101}-\tilde E_{100}-\tilde E_{001}+\tilde E_{000}$, which also depends on the anharmonicities and therefore cannot be nulled by the single control parameter $\Delta(t)$. Here, $\tilde E_{ijk}$ denote the eigenenergies of the computational basis states defined in Sec. \ref{sec:leak_2subspace}. Within the symmetric model considered here, the standard fourth-order perturbative expressions for $\zeta$ \cite{Sung2021PRX, Heunisch2023PRApplied} collapses to
\begin{equation}
\zeta \approx -\frac{4g_{12}^2}{\alpha} + g_{12}g^2\left(\frac{8}{\alpha\Delta_0}+\frac{4}{\Delta_0^2}\right) -\frac{4g^4}{\Delta_0^2\alpha}-\frac{8g^4}{\Delta_0^2(2\Delta_0+\alpha_c)} .
\end{equation}
Inserting $g_{12} = g^2/\Delta_0$ from Eq.~(\ref{eqn:Delta_0}), all terms containing the qubit anharmonicity cancel, and only a single term survives,
\begin{equation}
\zeta \approx \frac{4g^4\alpha_c}{\Delta_0^3\,(2\Delta_0+\alpha_c)}
      \simeq 2\alpha_c\left(\frac{g_{12}}{g}\right)^{4} \ .
\end{equation}
At the off-point the residual $ZZ$ interaction is thus governed solely by the coupler anharmonicity and the finite direct qubit--qubit coupling $g_{12}$.

\section{Correction in the Superadiabatic frame}
\label{sec:SAD_correction}

We show here how our general technique can be extended to higher orders in a manner that avoids analyzing higher-order terms in the Magnus expansion (something that can be unwieldy).  Instead, higher-order corrections can be obtained by working in a so-called ``superadiabatic" frame, instead of the more conventional adiabatic frame.  We illustrate this for the single-excitation leakage problem introduced in Sec.~\ref{sec:leak_corr}.

To effectively push what would have been higher-order terms in the Magnus expansion to leading order, we transform the problem to the superadiabatic frame.  This is the frame where the eigenstates of $H_{\text{ad}}$ (cf.~Eq. (\ref{eqn:H_ad_2x2})) become stationary. For simplicity, we first map the adiabatic states to the standard computational basis, i.e. $\ket{B_c} \equiv \ket{0}$ and $\ket{B_q} \equiv \ket{1}$. In this representation, the bare adiabatic Hamiltonian $H_{\text{ad}}(t)$ (Eq. (\ref{eqn:H_ad_2x2})) and the adiabatic control Hamiltonian $H_{\text{ctrl, ad}}(t)$ (Eq. (\ref{eqn:H_ad})) can be expressed via Pauli matrices as
\begin{align}
    H_{\text{ad}}(t) &= \frac{E_c - E_q}{2} \sigma_z + \dot{\theta} \sigma_y \\
    H_{\text{ctrl, ad}}(t) &= \frac{\delta \omega}{2} \bigl(\mathds{1} + \cos 2\theta \, \sigma_z + \sin 2\theta \, \sigma_x \bigr),
\end{align}
with $\mathds{1}$ being the identity operator and all coefficients are time-dependent functions.

We can move to the superadiabatic frame via the unitary transformation 
$V(t) = \exp\bigl(-i\nu(t) \sigma_x\bigr)$. The superadiabatic mixing angle $\nu(t)$, defined to eliminate the non-adiabatic $\sigma_y$-coupling present in the adiabatic Hamiltonian, is given by
\begin{equation}
    \tan 2\nu = -\frac{2\dot{\theta}}{E_c - E_q} \ .
\end{equation}
Applying this unitary transformation to the total adiabatic Hamiltonian $H_{\text{ad}}' = H_{\text{ad}} + H_{\text{ctrl, ad}}$ yields the total superadiabatic Hamiltonian $ H_{\text{sad}}' = V^\dagger \bigl(H_{\text{ad}} + H_{\text{ctrl, ad}}\bigr) V - i V^\dagger \dot{V} $, which is explicitly given by
\begin{align}
    H_{\text{sad}}' &= \frac{\tilde{E}_c - \tilde{E}_q}{2} \sigma_z + \frac{\delta \omega}{2} \Bigl(\mathds{1} + \sin 2\theta \, \sigma_x  \notag \\
    &+ \cos 2\theta \bigl( \cos 2\nu \, \sigma_z + \sin 2\nu \, \sigma_y \bigr) \Bigr) - \dot{\nu} \sigma_x \ .
\end{align}
Here, the modified superadiabatic eigenenergies are given by $\tilde{E}_{c,q} = \frac{1}{2}\bigl(E_c + E_q \pm \sqrt{(E_c-E_q)^2 + 4\dot{\theta}^2}\bigr)$, where the plus and minus signs correspond to the indices $c$ and $q$, respectively.

In the superadiabatic frame, the residual non-adiabatic error $\propto \dot{\nu}$ is now purely $\sigma_x$-like. However, our control Hamiltonian is more complex.  In addition to having the desired $\sigma_x$ component (which can be tuned to cancel the non-adiabatic error term), it also has components along $\sigma_z$ and $\sigma_y$.  
The $\sigma_z$ component can be ignored:  it only introduces a slight phase shift that can be compensated by a recalibration of the coupler hold detuning $\Delta_h$. The $\sigma_y$ term appears problematic, but can be 
accounted for using our general integration-by-parts strategy, as we now show. 

As in Sec.~\ref{sec:leak_corr}, we wish to find a pulse correction 
$\delta \omega(t)$ that causes the first order Magnus Hamiltonian to vanish. We can write the corresponding condition in a manner analogous to Eq. (\ref{eqn:Null_cond}), finding:
\begin{align}
    0 \overset{!}{=} \int_0^{t_{g}} &\biggl(-\dot{\nu} + \frac{\delta\omega}{2} \sin 2\theta \notag \\
    &- i \frac{\delta\omega}{2} \cos 2\theta \sin 2\nu \, \biggr)e^{i\tilde{\Phi}(t)} \, dt \ ,
\end{align}
where $\tilde{\Phi}(t) = \int_0^t (\tilde{E}_c - \tilde{E}_q) \, dt'$. 
We have a version of the same problem that arose in the main text: the correction $\delta \omega(t)$ is not in phase with the error $\propto \dot{\nu}$, meaning we cannot simply find $\delta \omega(t)$ by setting the integrand to zero.  The solution is to again use integration by parts.  Here, however, we 
do this in a way to make the entire integrand imaginary.  This has the virtue of leading to a simpler correction which does not involve higher-order pulse derivatives.  
Doing this yields the defining differential equation for the correction pulse $\delta\omega(t)$:
\begin{equation}
    \frac{d}{dt} \left( \frac{\delta\omega \sin 2\theta-2\dot{\nu}}{\tilde{E}_c - \tilde{E}_q} \right) = \delta\omega \cos 2\theta \sin 2\nu \ .
    \label{eqn:ODE_SAD}
\end{equation}

To solve this differential equation analytically, we expand the time derivative on the left-hand side using the product rule. Substituting the geometric relation $\sin 2\nu = -2\dot{\theta} / (\tilde{E}_c - \tilde{E}_q)$ into the right-hand side, the equation becomes a linear first-order ordinary differential equation for $\delta\omega$,
\begin{equation}
    \sin 2\theta \, \frac{d}{dt} \left( \frac{\delta\omega}{\tilde{E}_c -
    \tilde{E}_q} \right) + \frac{\delta\omega \cdot 4 \dot{\theta} \cos 2\theta}
    {\tilde{E}_c - \tilde{E}_q} = \frac{d}{dt} \left( \frac{2\dot{\nu}}
    {\tilde{E}_c - \tilde{E}_q} \right) \,.
\end{equation}
Multiplying by $\sin 2\theta$ and using $\tfrac{d}{dt} \sin^2 2\theta = 4 \dot{\theta} \sin 2\theta \cos 2\theta$, the left-hand side collapses into the total time derivative of $\delta\omega \sin^2 2\theta / (\tilde{E}_c - \tilde{E}_q)$. A single integration then isolates the correction pulse,
\begin{align}
    \frac{\delta\omega(t)}{\tilde{E}_c - \tilde{E}_q} &= \frac{1}{\sin^2 2\theta(t)} \notag \\
    &\times \int_0^t \sin 2\theta(t_1) \frac{d}{dt_1} \left( \frac{2\dot{\nu}(t_1)}{\tilde{E}_c(t_1) - \tilde{E}_q(t_1)} \right) dt_1 \ .
\end{align}
To avoid calculating $\ddot{\nu}$, which depends on the third time derivative of $\theta$, we apply integration by parts to shift the derivative away from $\dot{\nu}$ and onto the trigonometric prefactor. Assuming a $\mathcal{C}^2$-smooth pulse, the exact analytical solution simplifies to
\begin{equation}
\label{eqn:SAD_corr}
    \delta\omega(t) = \frac{2\dot{\nu}}{\sin 2\theta} - \frac{\tilde{E}_c -
    \tilde{E}_q}{\sin^2 2\theta} \int_0^t \frac{4 \dot{\nu} \dot{\theta}
    \cos 2\theta}{\tilde{E}_c - \tilde{E}_q} \, dt_1 \,.
\end{equation}
This exact final result for the correction pulse $\delta \omega(t)$ has a number of advantageous features.  Note that while it depends on $\dot{\nu}$, it does not depend on higher derivatives of this function. Since $\nu \propto \dot{\Delta}$, the superadiabatic leakage suppression entirely avoids the need for higher-order pulse derivatives, meaning that the smoothness constraint on the base pulse is the same as before, i.e.~one needs a 
standard $\mathcal{C}^2$-smooth base pulse.
\begin{figure}[!htp]
    \centering
    \includegraphics[width=\columnwidth]{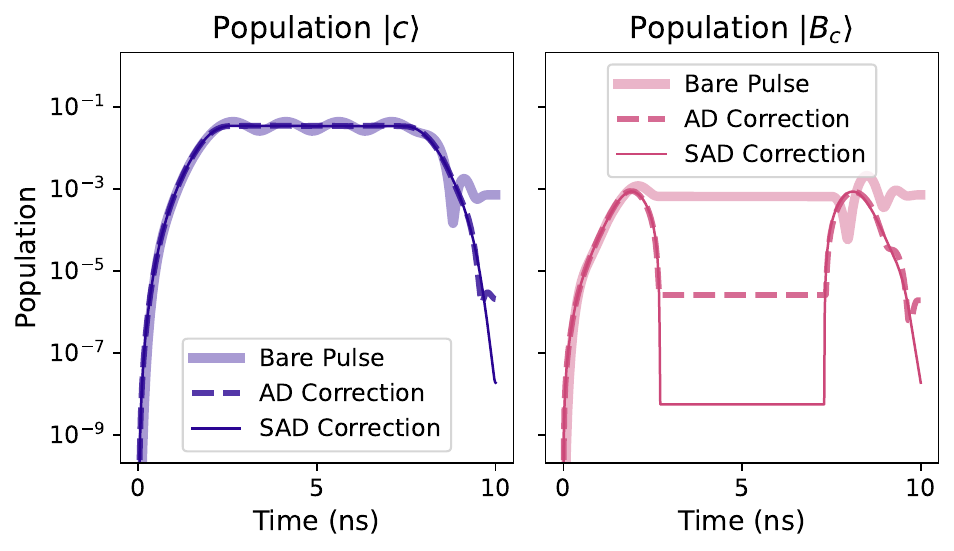}
    \caption{Population of the bare coupler state $\ket{c}$ (left) and the parasitic coupler-like bright state $\ket{B_c}$ (right) during a $10\,\rm{ns}$ iSWAP gate. While the first-order correction derived in the adiabatic frame (cf. Fig. \ref{fig:corr1}) already significantly reduces final leakage by forcing the mean evolution on the adiabatic trajectory, this plot demonstrates the enhanced performance of the next-order correction. By applying the pulse from Eq. (\ref{eqn:SAD_corr}), leakage is suppressed by nearly three more orders of magnitude compared to the correction derived in the adiabatic frame as the system is strictly forced onto the superadiabatic trajectory. Simulation parameters are detailed in Table \ref{tab:par_iSWAP}.}
    \label{fig:corr_sad}
\end{figure}

As illustrated in Fig.~\ref{fig:corr_sad}, the correction derived in the superadiabatic frame suppresses the residual populations of $\ket{B_c}$ and $\ket{c}$ in the lab-frame by nearly three orders of magnitude compared to the adiabatic approach. This drastic improvement is rooted in the fundamental scaling of the residual errors. In the adiabatic frame, the leading uncorrected second-order Magnus error scales as $\int dt_1 \int_0^{t_1} dt_2 \, \dot{\theta}(t_1) \dot{\theta}(t_2) \mathcal{P}$, where $\mathcal{P}$ represents a dimensionless Pauli operator. In the superadiabatic frame, however, the equivalent residual error scales with $\int dt_1 \int_0^{t_1} dt_2 \, \dot{\nu}(t_1) \dot{\nu}(t_2) \mathcal{P}$. Introducing the dispersive parameter $\eta = g/\Delta$ and the adiabaticity parameter $\epsilon = \dot{\Delta}/\Delta^2$, we find that the adiabatic drive scales as $\dot{\theta} \sim \Delta \eta \epsilon$. In contrast, the superadiabatic drive scales as $\dot{\nu} \propto \ddot{\theta}/\Delta \sim \Delta \eta \epsilon^2$. One might naively argue that, given a smooth $\mathcal{C}^2$ base pulse, this favorable $\epsilon^2$ scaling could also be achieved purely within the adiabatic frame by repeatedly applying integration by parts to the bare $\dot{\theta}$ error \cite{Lidar2009JMP, Rezakhani2010PRA}. However, this strategy ultimately fails because the required control field $\delta\omega \sim \ddot \theta$ intrinsically induces crosstalk. Since $\delta\omega$ must dynamically counteract the bare error, its crosstalk operates on the same effective magnitude as $\dot{\theta}$ itself. This dynamic crosstalk cannot simply be integrated away without generating uncorrectable higher-order pulse derivatives. Therefore, the core strategy of the superadiabatic frame is not to laboriously evaluate higher-order Magnus terms, but to geometrically suppress the amplitude of the error terms before the perturbative expansion is even applied.

\section{Magnitude of the Correction Pulse}
\label{sec:Convergence}
In this section we want to demonstrate in which regimes Eq. (\ref{eqn:corr1}) is small as required for a leading-order Magnus-based approach. From Eq. (\ref{eq:theta}) one has the exact relation
\begin{equation}
  \sin 2\theta = - \frac{2\sqrt{2}\,g}{E_c - E_q} \ ,
  \label{eqn:app_sin}
\end{equation}
so that the transverse projection of the longitudinal control is given by the ratio of the coupling to the instantaneous gap. Differentiating Eq. (\ref{eq:theta}) and using $\sec^{2}2\theta = (E_c-E_q)^{2}/(\Delta-g_{12})^{2}$ yields the corresponding
expression for the diabatic drive,
\begin{equation}
  \dot\theta = \frac{\sqrt{2}\,g\,\dot\Delta}{(E_c-E_q)^{2}}  ,
  \label{eqn:app_thetadot}
\end{equation}
in which the gap enters algebraically and need not be differentiated. This is no longer the case for the correction itself, whose time derivative requires calculating
\begin{equation}
  \frac{d}{dt}\big(E_c-E_q\big) = \frac{(\Delta-g_{12})\,\dot\Delta}{E_c-E_q} \ .
  \label{eqn:app_gapdot}
\end{equation}
Combining Eq. (\ref{eqn:app_thetadot}) and (\ref{eqn:app_gapdot}) we find
\begin{equation}
  \frac{d}{dt}\frac{\dot\theta}{E_c-E_q} = \sqrt{2}\,g
  \left(\frac{\ddot\Delta}{(E_c-E_q)^{3}}
  - \frac{3(\Delta-g_{12})\,\dot\Delta^{2}}{(E_c-E_q)^{5}}\right) \ .
\end{equation}
Inserting this result and the geometric prefactor from Eq. (\ref{eqn:app_sin}) into Eq. (\ref{eqn:corr1}) we find
\begin{equation}
  \delta\omega = \frac{\ddot\Delta}{(E_c-E_q)^{2}}
  - \frac{3\,(\Delta-g_{12})\,\dot\Delta^{2}}{(E_c-E_q)^{4}} \,.
  \label{eqn:app_closed}
\end{equation}
The correction is therefore fixed by the base pulse and its first two derivatives alone. Since $E_c - E_q \geq 2\sqrt{2}\,g$ for any detuning (cf. Eq. (\ref{eqn:Eq}) and Eq. (\ref{eqn:Ec})), the gap is bounded from below by a circuit parameter rather than by a pulse envelope, and the only singular limit of Eq. (\ref{eqn:app_closed}) is $g \to 0$, where the coupler decouples from the qubits and no gate is realized. Measured against the gap that generates the dynamical phase $\Phi(t)$, the correction remains small. For the pulse of Fig. \ref{fig:corr1} we find
$\max|\delta\omega|/(E_c-E_q) \approx 0.039$, so that it acts only as a small perturbation of the base pulse. 

The same closed form also settles the limit $g_{12} \to 0$, in which the off-point Eq. (\ref{eqn:Delta_0}) is pushed to $\Delta_0 \to \infty$ and $\sin 2\theta_0 \to 0$, such that the prefactor of Eq. (\ref{eqn:corr1}) appears to diverge at $t=0$ and $t=t_g$. No such divergence occurs in Eq. (\ref{eqn:app_closed}), where the mixing angle has cancelled and the gap, which does not close, enters the denominators
only.

\section{Numerically exact evaluation of the correction pulse}
\label{sec:numerics}
The derivation in Sec.~\ref{sec:leak_2subspace} relies on a perturbative Schrieffer--Wolff transformation to obtain the adiabatic frame. In practice we evaluate the same nulling condition using numerically exact instantaneous eigenstates, which removes the reliance on the perturbative expansion. In the bare basis $\{|101\rangle, |110\rangle, |011\rangle, |200\rangle, |002\rangle, |020\rangle\}$ the full two-excitation block of Eq.~(\ref{eqn:Hardware_H}) reads
\begin{equation}
H_2(t)=
\begin{pmatrix}
0 & g & g & \sqrt2 g_{12} & \sqrt2 g_{12} & 0\\
g & \Delta(t) & g_{12} & \sqrt2 g & 0 & \sqrt2 g\\
g & g_{12} & \Delta(t) & 0 & \sqrt2 g & \sqrt2 g\\
\sqrt2 g_{12} & \sqrt2 g & 0 & \alpha & 0 & 0\\
\sqrt2 g_{12} & 0 & \sqrt2 g & 0 & \alpha & 0\\
0 & \sqrt2 g & \sqrt2 g & 0 & 0 & 2\Delta(t)+\alpha_c
\end{pmatrix} 
\label{eqn:full2ndexcspace}
\end{equation}
which is the exact counterpart of the truncated model of Eq.~(\ref{eqn:2excH}): it retains the doubly excited coupler state $|020\rangle$ as well as the antisymmetric combinations, so that no state symmetry is assumed. Note that transforming Eq.~(\ref{eqn:full2ndexcspace}) into the basis of Eq.~(\ref{eqn:2excH}) shifts the $|C_+\rangle$ state by $+g_{12}$. As this is only a
small constant shift, we absorb it into $\Delta(t)$ in
Eq.~(\ref{eqn:2excH}).

We diagonalize $H_2(t)$ on the time grid of the pulse. The resulting eigenvectors are the columns of the exact transformation $U(t)$ into the adiabatic frame, i.e. the numerically exact analogue of $\exp(\hat S)$ with $\hat S$ from Eq.~(\ref{eqn:S_generator}). $\dot U$ is then obtained by numerical differentiation. Since $H_2(t)$ is real symmetric, its eigenvectors can be chosen real. Only a discrete sign gauge freedom remains, which we track at each time step throughout our calculation.

The control Hamiltonian is a modulation of the coupler frequency only, so within this subspace it is proportional to the coupler excitation number, $H_{\text{ctrl}}(t)=\delta\omega(t)\,\hat n_c$ with $\hat n_c=\mathrm{diag}(0,1,1,0,0,2)$ in the above basis. The diabatic couplings are the matrix elements of the inertial term $-i\dot U U^\dagger$. Evaluating the leading-order Magnus condition of Eq.~(\ref{eqn:Null_cond2}) with these exact quantities, and applying the same integration by parts that brings the diabatic term in phase with the real-valued control, yields
\begin{equation}
\delta\omega(t)=\frac{1}{\langle\tilde{101}|\hat n_c|\tilde Q_+\rangle}\,\frac{d}{dt}\!\left(\frac{\langle\tilde{101}|\partial_t|\tilde Q_+\rangle}{\tilde E_{Q_+}(t)-\tilde E_{101}(t)}\right) \ .
\label{eqn:corr_numexact}
\end{equation}
Equation~(\ref{eqn:corr_numexact}) is the form used for all simulations in Secs.~\ref{sec:Hardware_result},  \ref{sec:two_channel_corr} and \ref{sec:Slepian}. The secondary
correction $\delta\omega^{(2)}$ is obtained in the same way from the corresponding matrix elements of $|\tilde C_+\rangle$.

\section{Robustness against Parameter Asymmetries}
\label{sec:robustness}
\begin{figure}[!htp]
    \centering
    \includegraphics[width = \columnwidth]{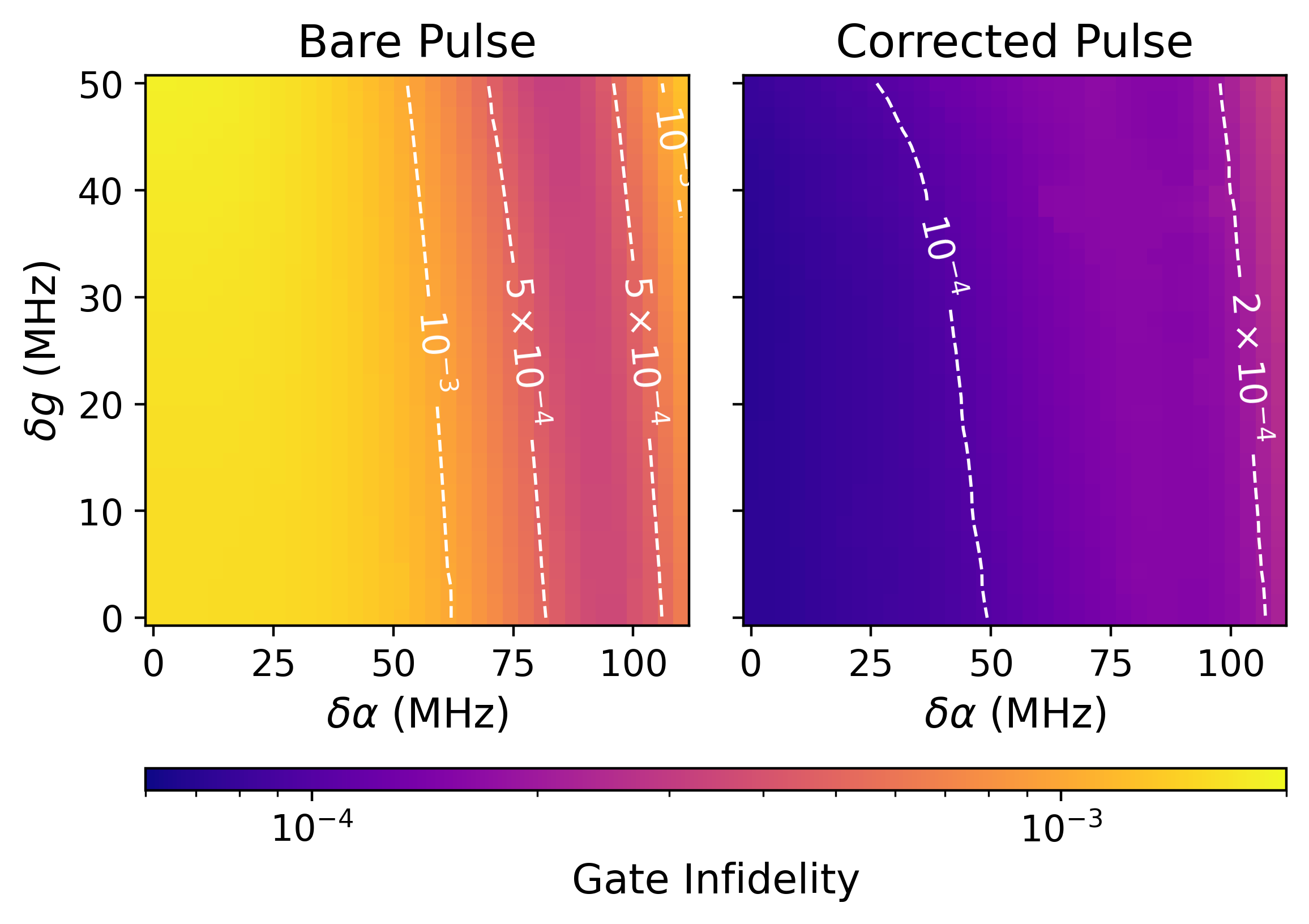}
    \caption{Robustness of the iSWAP gate fidelity against parameter asymmetries. The fidelity of a 20\,ns gate with a fixed ramp time of 6\,ns is plotted as a function of the anharmonicity difference $\delta \alpha = \alpha_1 - \alpha_2$ and coupling asymmetry $\delta g = g_1 - g_2$. Despite being derived specifically to suppress the dominant $\ket{Q_+}$ leakage state, the correction pulse remains highly effective, yielding significant fidelity improvements across a broad parameter regime. Both panels share a common color scale, contour labels indicate the gate infidelity. Simulation parameters are detailed in Table \ref{tab:par_iSWAP}.}
    \label{fig:robustness}
\end{figure}
In this section, we analyze the robustness of our correction scheme against system asymmetries, specifically when the qubits exhibit unequal anharmonicities ($\alpha_1 \neq \alpha_2$) or asymmetric coupling strengths to the tunable coupler ($g_1 \neq g_2$). Under these conditions, the effective three-level approximation in Eq. (\ref{eqn:2excH}) within the two-excitation subspace breaks down. Consequently, the asymmetric states $\ket{C_-} = \frac{1}{\sqrt{2}}(\ket{110} - \ket{011})$ and $\ket{Q_-} = \frac{1}{\sqrt{2}}(\ket{200} - \ket{002})$ become visible to the computational state $\ket{101}$. Despite this added complexity, applying the correction pulse that was originally derived to solely suppress leakage into the symmetric state $\ket{\tilde{Q}_+}_0$ still yields a remarkable fidelity improvement across a broad parameter regime, as illustrated in Fig. \ref{fig:robustness}.

The sustained performance of the correction pulse can be understood by examining the distinct physical effects of each asymmetry. For anharmonicity differences $\delta\alpha = \alpha_1 - \alpha_2$, the antisymmetric state $\ket{Q_-}$ couples to $\ket{Q_+}$ with coupling strength $\delta \alpha$ but only connects to $\ket{101}$ strictly via a third-order process. As a result, its parasitic population remains naturally suppressed. 

In contrast, the leading-order consequence of using unequal coupling strengths is a residual SWAP error originating from unequal Stark shifts $\sim g_i^2/\Delta$ of the qubit frequencies. These shifts break the strict resonance condition between the qubits, which leaves a minor residual population in the initial state and introduces an error. In practice, however, this effect can be easily mitigated by slightly recalibrating the transmon frequencies via an external flux pulse. For this reason, the results shown in Fig. \ref{fig:robustness} were obtained through an optimization that includes not only the hold frequency $\Delta_h$ of the tunable coupler but also small variations of the qubit frequencies to compensate for the SWAP error. With this compensation, the residual leakage into the antisymmetric state $\ket{\tilde{C}_-}_0$ arising from the asymmetric coupling and into the state $\ket{\tilde{Q}_-}_0$ via a second-order process becomes negligible. The overall error budget therefore remains limited by population leakage into the state $\ket{\tilde{Q}_+}_0$, demonstrating the overall effectiveness of our correction.

\section{CZ gate}
\label{sec:CZ_correction}
\begin{figure}[!htp]
    \centering
    \includegraphics[width=\columnwidth]{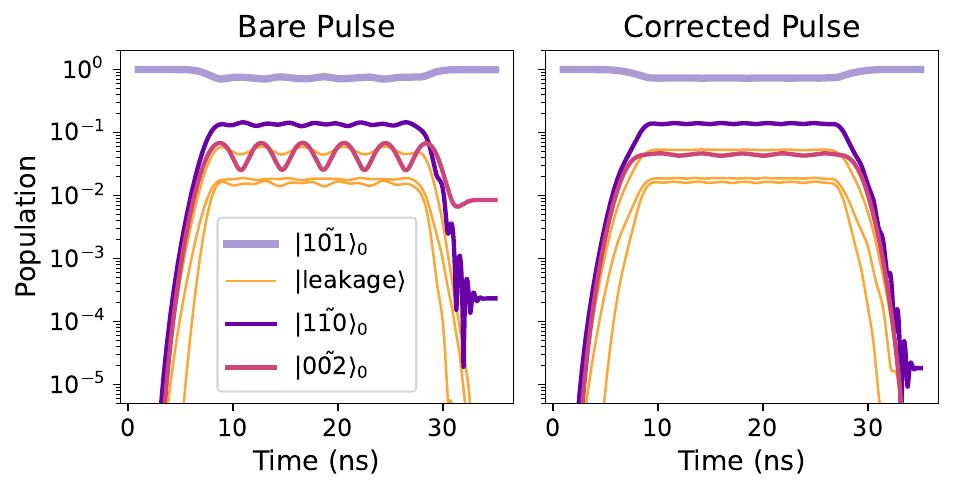}
    \caption{State populations in the second-excited subspace for the CZ gate simulation. Using just the bare pulse for the coupler (left), significant leakage remains in the $\ket{\tilde{002}}_0$ state, limiting the overall gate fidelity. Applying the analytical correction suppresses the leakage population of this state, reducing the gate infidelity by two orders of magnitude (right). The simulation parameters are listed in Table \ref{tab:par_CZ}.}
    \label{fig:corr_CZ}
\end{figure}
In this section, we demonstrate that our leakage elimination mechanism can be readily adapted to a diverse range of baseband-flux-activated gates. A prominent example is the controlled-phase (CZ) gate. Crucially, in this specific implementation scheme we are utilizing a flux-tunable coupler to activate conditional phase accumulation via a strong $ZZ$-interaction as proposed in \cite{Collodo2020PRL}. Unlike the symmetric iSWAP gate, this system entirely lacks spatial and state symmetries due to the use of detuned qubits with bare frequencies of $\omega_1/2\pi = 4.75\,\rm{GHz}$ and $\omega_2/2\pi = 5.2\,\rm{GHz}$.

As illustrated on the left side of Fig. \ref{fig:corr_CZ}, an uncorrected flux pulse induces severe leakage into the non-computational $\ket{\tilde{002}}_0$ state, as the coupler drives the system close to a resonance within the crowded two-excitation subspace. By applying the analytical framework described in Section \ref{sec:leak_2subspace}, we analogously derive a tailored correction pulse that successfully suppresses residual leakage at the gate's end by several orders of magnitude. This mitigation dramatically drops the gate infidelity for a $35\,$ns CZ gate from $1-f = 2.2\times 10^{-3}$ without correction to $1-f = 1.9\times 10^{-5}$ with the active correction, highlighting the exceptional versatility of our analytical pulse engineering scheme across varied and highly asymmetric architectural settings.

\begin{table}[htpb]
    \centering
    \setlength{\tabcolsep}{2.0em} 
    \renewcommand{\arraystretch}{1.25} 
    \begin{tabular}{cc}
        \toprule
        \textbf{Parameters} &  \\
        \midrule
        $\alpha_1/2\pi$ & $-240\,\mathrm{MHz}$ \\
        $\alpha_2/2\pi$ & $-252\,\mathrm{MHz}$ \\
        $\alpha_c/2\pi$ & $-187\,\mathrm{MHz}$ \\
        $\omega_1/2\pi$ & $4.75\,\mathrm{GHz}$ \\
        $\omega_2/2\pi$ & $5.2\,\mathrm{GHz}$ \\
        $g/2\pi$ & $195\,\mathrm{MHz}$ \\
        $g_{12}/2\pi$ & $22.3\,\mathrm{MHz}$ \\
        \bottomrule
    \end{tabular}
    \caption{System parameters used for the numerical simulation of the CZ gate based on Eq. (\ref{eqn:Hardware_H}) in Fig. \ref{fig:corr_CZ}.}
    \label{tab:par_CZ}
\end{table}

\section{Comparison with $\Phi$-DRAG \cite{Georgiadis2026arxiv}}
\label{sec:PhiDRAG}
Recently, Georgiadis \textit{et al.}~\cite{Georgiadis2026arxiv} suggested a method to suppress leakage in flux-activated gates as well. While their approach relies on suppressing the spectral weight at a fixed leakage transition frequency, we employ a more general Magnus strategy~\cite{Ribeiro2017PRX} that does not identify the leakage integral in Eq.~\eqref{eqn:Null_cond2} with a simple Fourier ansatz. Instead, we integrate by parts directly to treat the time dependence of the instantaneous gap $\Phi_{13}(t)$ and its time derivative explicitly. This gives us greater flexibility regarding the leakage states we wish to suppress, as we are not tied to a single, fixed transition frequency by retaining the full time dependence of the energy gap.

This time dependence can be small in some scenarios, for example, when the gap associated with the leakage transition to $\ket{\tilde{Q}_+}$ is set by the anharmonicity and acquires a time dependence only at second order in $g$, as described in Sec.~\ref{sec:leak_2subspace}. In this case, we can directly benchmark our approach against theirs.

\begin{figure}
    \centering
    \includegraphics[width = \linewidth]{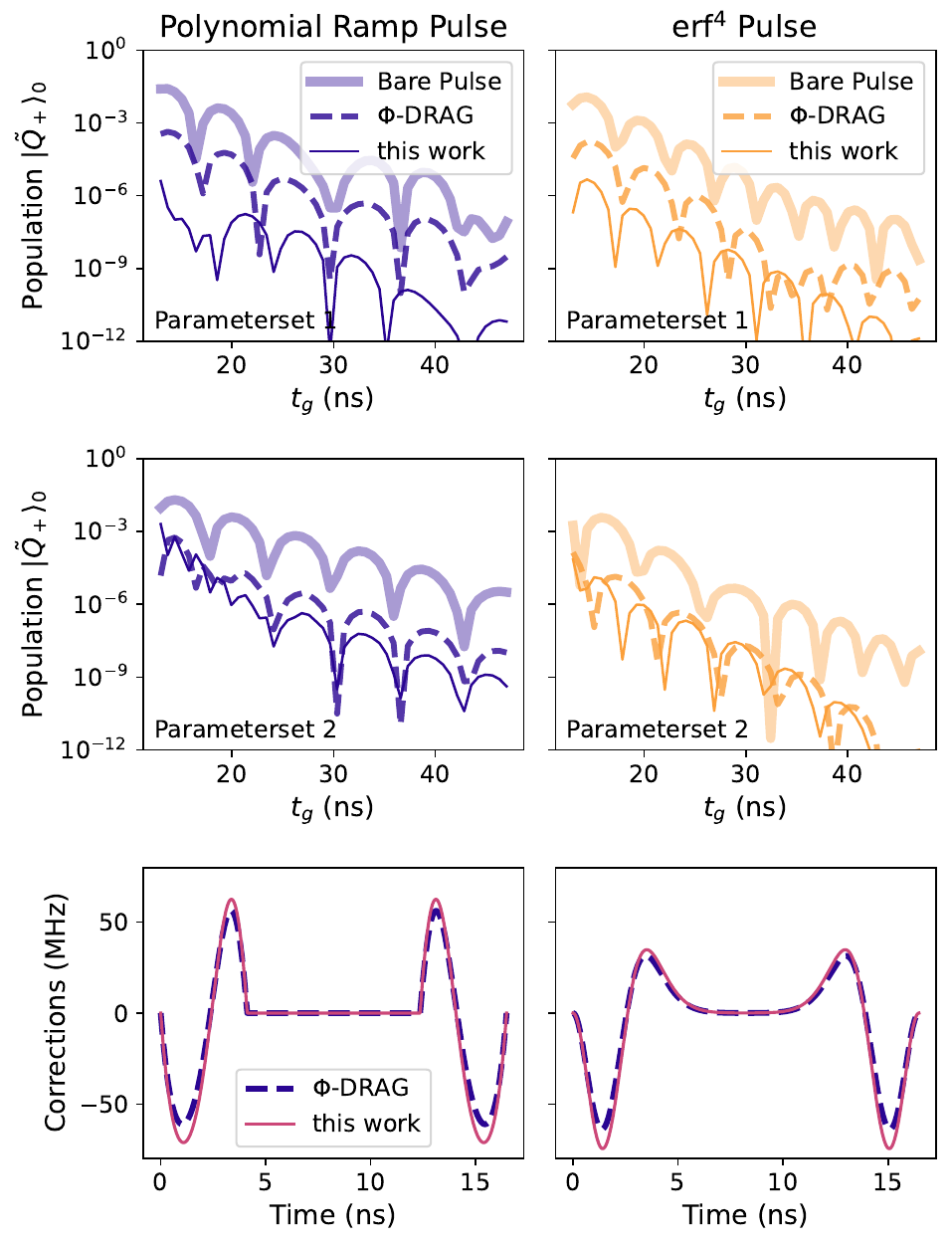}
    \caption{Comparison of the correction derived in Sec. \ref{sec:leak_2subspace} with the spectral correction of Ref. \cite{Georgiadis2026arxiv}, simulated in the reduced two-excitation subspace of Eq. (\ref{eqn:2excH}). Top and middle rows show the residual population of the leakage state $\ket{\tilde Q_+}_0$ at the end of the gate as a function of the gate time $t_g$, for the two parameter sets of Table \ref{tab:parPhiDrag}. The left column uses the polynomial ramp of Eq. (\ref{eqn:Poly}), the right column the erf$^4$ profile. The corrections explicitly shown in the bottom panels are derived for a $16.5\,\rm{ns}$ gate using parameter set 1.}
    \label{fig:phidrag}
\end{figure}

In Fig.~\ref{fig:phidrag} we compare the ability of both methods to suppress leakage into the $\ket{\tilde Q_+}_0$ state. To this end we simulate the reduced Hamiltonian of Eq.~(\ref{eqn:2excH}) for two pulse shapes: the polynomial ramp of Eq.~(\ref{eqn:Poly}), and a profile built from error functions,
\begin{align}
    \Delta (t) = \Delta_0 + (\Delta_h - \Delta_0) \, \text{erf}^4 \left( \frac{t}{\tau_r} \right) \, \text{erf}^4 \left( \frac{t_g-t}{\tau_r} \right)
\end{align}
which is the flux profile employed in \cite{Georgiadis2026arxiv}, here applied directly to the coupler detuning. Both shapes are evaluated for two parameter sets differing in the dispersive ratio $g/\Delta$ (see Table~\ref{tab:parPhiDrag}). In all cases the hold detuning $\Delta_h$ is fixed by the same gate condition in Eq.~(\ref{eqn:gate_cond}).

Although the ansatz in Ref.~\cite{Georgiadis2026arxiv} is simpler, it yields the correction in terms of a modulated effective coupling,
\begin{equation}
    g_{\mathrm{eff}}(t) \to g_{\mathrm{eff,corr}}(t) = g_{\mathrm{eff}}(t) + \frac{1}{\alpha^2} \frac{d^2}{dt^2} g_{\mathrm{eff}}(t),
\end{equation}
rather than directly in terms of the pulse detuning $\Delta(t)$. To ensure a fair comparison we do not use the perturbative Schrieffer--Wolff expression for the effective coupling, but extract $g_\mathrm{eff}$ by exact block diagonalization of Eq.~(\ref{eqn:2excH}). The correction of \cite{Georgiadis2026arxiv} is then evaluated from this coupling and mapped back onto a detuning profile $\Delta(t)$ by numerically inverting
$g_\mathrm{eff}(\Delta)$.

As shown in the lower panels of Fig.~\ref{fig:phidrag}, the two corrections have a very similar shape, reflecting the common fundamental structure of both correction strategies. For the more dispersive parameter set (middle panels), where the gap of the $\ket{\tilde Q_+}_0$ transition is close to the bare anharmonicity, the two corrections yield a comparable improvement over the bare pulse. For the larger dispersive ratio (top panels) the transition frequency can no longer be treated as constant over the pulse, and our correction suppresses this leakage channel by another order of magnitude.

\begin{table}[htpb]
    \centering
    \setlength{\tabcolsep}{2.0em}
    \renewcommand{\arraystretch}{1.25}
    \begin{tabular}{ccc}
        \toprule
        \textbf{Parameters} & \textbf{Set 1} & \textbf{Set 2} \\
        \midrule
        $g/2\pi$ & $210\,\mathrm{MHz}$ & $90\,\mathrm{MHz}$ \\
        $g_{12}/2\pi$ & $36\,\mathrm{MHz}$ & $6\,\mathrm{MHz}$ \\
        $\alpha/2\pi$ & $-270\,\mathrm{MHz}$ & $-270\,\mathrm{MHz}$ \\
        $\Delta_0/2\pi$ & $1.189\,\mathrm{GHz}$ & $1.344\,\mathrm{GHz}$ \\
        $\tau/t_g$ & $0.25$ & $0.25$ \\
        $\tau_r/t_g$ & $0.15$ & $0.15$ \\
        \bottomrule
    \end{tabular}
    \caption{System parameters used for the numerical simulation of Eq. (\ref{eqn:2excH}) in Fig. \ref{fig:phidrag}. The off-point $\Delta_0$ follows from Eq. (\ref{eqn:Delta_0}), $\tau$ denotes the ramp time of the polynomial pulse and $\tau_r$ the rise time of the erf$^4$ profile.}
    \label{tab:parPhiDrag}
\end{table}